\documentclass[useAMS,usenatbib]{mnras}
\usepackage[pdftex]{graphicx}
\usepackage{amsmath}
\usepackage{amsfonts}
\usepackage{amssymb}
\usepackage{subcaption}
\usepackage{mwe}

\def\q#1{`#1'}

\newcommand{\ap}{$\sim$ }
\newcommand{\ts}{\textsuperscript}
\title[An SMA view of the dust ridge]{Star formation in a high-pressure environment:\\ An SMA view of the Galactic centre dust ridge}
\author[D. L. Walker et al.]{D.~L.~Walker$^{1,2,3,9}$\thanks{E-mail: \texttt{daniel.walker@alma.cl}},
S.~N.~Longmore$^{3}$,
Q. Zhang$^{4}$,
C. Battersby$^{4,5}$,
E. Keto$^{4}$,
\and J.~M.~D.~Kruijssen$^{6,7}$,
A. Ginsburg$^{8}$,
X. Lu$^{9}$, 
J.~D.~Henshaw$^{3,7}$,
J. Kauffmann$^{10,11}$,
\and T. Pillai$^{10}$,
E. A. C. Mills$^{12}$,
A. J. Walsh$^{13}$,
J. Bally$^{14}$,
L. C. Ho$^{15,16}$,
K. Immer$^{17}$
and 
\and K. G. Johnston$^{18}$ \vspace{0.2cm} \\
$^{1}$Joint ALMA Observatory, Alonso de Cordova 3107, Vitacura,
Santiago, Chile\\
$^{2}$National Astronomical Observatory of Japan, Alonso de Cordova 3788, 61B Vitacura, Santiago, Chile\\
$^{3}$Astrophysics Research Institute, Liverpool John Moores University, IC2, 146 Brownlow Hill, Liverpool, L3 5RF, United Kingdom\\
$^{4}$Harvard-Smithsonian Center for Astrophysics, 60 Garden Street, Cambridge, MA 02138, USA\\
$^{5}$University of Connecticut, Department of Physics, 2152 Hillside Road, U-3046, Storrs, CT 06269, USA\\
$^6$Astronomisches Rechen-Institut, Zentrum f\"{u}r Astronomie der Universit\"{a}t Heidelberg, M\"{o}nchhofstra\ss e 12-14, 69120 Heidelberg, Germany\\
$^7$Max-Planck Institut f\"{u}r Astronomie, K\"{o}nigstuhl 17, 69117 Heidelberg, Germany\\
$^8$National Radio Astronomy Observatory, P.O. Box O 1009, Lopezville Drive, Socorro, NM 87801, USA\\
$^9$National Astronomical Observatory of Japan, 2-21-1 Osawa, Mitaka,Tokyo, 181-8588, Japan\\
$^{10}$Max-Planck-Institut f\"{u}r Radioastronomie, Auf dem H\"{u}gel 69, 53121 Bonn, Germany\\
$^{11}$Haystack Observatory, Massachusetts Institute of Technology, 99 Millstone Road, Westford, MA 01886, USA\\
$^{12}$Department of Astronomy, Boston University, 725 Commonwealth Ave, Boston, MA 02215, USA\\
$^{13}$International Centre for Radio Astronomy Research, Curtin University, GPO Box U1987, Perth WA 6845, Australia\\
$^{14}$Department of Astrophysical and Planetary Sciences, University of Colorado, UCB 389, Boulder, CO 80309, USA\\
$^{15}$Kavli Institute for Astronomy and Astrophysics, Peking University, Beijing 100871, China\\
$^{16}$Department of Astronomy, School of Physics, Peking University, Beijing 100871, China\\
$^{17}$European Southern Observatory, Karl-Schwarzschild-Stra\ss e 2, D-85748 Garching bei M\"{u}nchen, Germany\\
$^{18}$School of Physics \& Astronomy, E.C. Stoner Building, The University of Leeds, Leeds, LS2 9JT, UK\\
}
\begin{document}
\date{}
\pagerange{\pageref{firstpage}--\pageref{lastpage}} \pubyear{2017}
\maketitle
\label{firstpage}
\begin{abstract}

The star formation rate in the Central Molecular Zone (CMZ) is an order of magnitude lower than predicted according to star formation relations that have been calibrated in the disc of our own and nearby galaxies. Understanding how and why star formation appears to be different in this region is crucial if we are to understand the environmental dependence of the star formation process. Here, we present the detection of a sample of high-mass cores in the CMZ's \q{dust ridge} that have been discovered with the Submillimeter Array. These cores range in mass from \ap 50 -- 2150~M$_{\odot}$ within radii of 0.1 -- 0.25~pc. {\textcolor{black}{All appear to be young}} (pre-UCH{\scriptsize II}), meaning that they are prime candidates for representing the initial conditions of high-mass stars and sub-clusters. We report that at least two of these cores (\q{c1} and \q{e1}) contain young, high-mass protostars. We compare all of the detected cores with high-mass cores and clouds in the Galactic disc and find that they are broadly similar in terms of their masses and sizes, despite being subjected to external pressures that are several orders of magnitude greater -- \ap 10$^{8}$~K~cm$^{-3}$, as opposed to \ap 10$^{5}$~K~cm$^{-3}$. The fact that \textgreater \ 80\% of these cores do not show any signs of star-forming activity in such a high-pressure environment leads us to conclude that this is further evidence for an increased critical density threshold for star formation in the CMZ due to turbulence.	
\end{abstract}


\newpage
\
\section{Introduction}\

Star formation relations provide the foundation for many astrophysical studies, from local to cosmological scales. As such, an end-to-end understanding of the star formation process is crucial if we are to apply it to the vastly differing conditions found throughout the Universe.

Empirical star formation relations have largely been calibrated using detailed studies of star forming regions in the disc of our own and nearby galaxies. Using observations of nearby star-forming regions, \citet{Lada10} proposed a gas surface density threshold for star formation of \ap 120~M$_{\odot}$~pc$^{-2}$, above which stars could form efficiently. This surface density threshold was argued to reflect an underlying volume density threshold of \ap 10$^{4}$~cm$^{-3}$. However, the regions from which relations like this are drawn are typically very similar to each other and do not probe significantly different environmental conditions, where such relations may not hold.

The conditions in the Galactic centre are known to be extreme compared to the Solar neighbourhood, with densities, gas temperatures, pressures, cosmic ray ionisation rates and magnetic field strengths being several factors to orders of magnitude greater \citep{Diederik_highz}. It therefore seems plausible that the process of star formation may proceed differently there. As it is our nearest \q{extreme} region, we can study the Galactic centre in great detail to investigate any potential environmental dependence of star formation.
 
It is already known that there is something different about the criteria under which stars form at the Galactic centre. \citet{snl_sf} showed that a significant fraction of the gas in the Central Molecular Zone (CMZ; inner \ap 500~pc of the Galaxy, Morris \& Serabyn 1996) lies above a volume density of \ap 10$^{4}$~cm$^{-3}$ -- the threshold proposed by \citet{Lada10}. Despite this, they found that the star formation rate (SFR) in the CMZ is 1 -- 2 orders of magnitude lower than predicted \citep[see][for a detailed study of star formation rates in the CMZ]{Barnes17}. The current understanding of this discrepant SFR is that the CMZ undergoes an episodic cycle and is currently at a low point due to the high turbulent energy density there \citep{Diederik_cmz_sf}, as evidenced by the comparatively large line-widths of \ap 10 -- 20~km s$^{-1}$ seen in the gas in the CMZ \citep{Henshaw_cmz}. This high turbulence will act to drive up the critical volume density threshold for star formation \citep{krumholz05, Padoan11}. Recent ALMA and SMA observations of the CMZ molecular cloud G0.253+0.016 show that the internal density structure of the cloud is consistent with a log-normal distribution that is well described by turbulent cloud models \citep{Brick_KJ, Jill_pdf_2014}.

Despite the seemingly low SFR in the CMZ, there are regions of substantial star formation and stellar content. Indeed, the star forming Sagittarius B2 complex is one of the most active sites of (high-mass) star formation in the Galaxy, hosting two dense clusters of H{\scriptsize II} regions \citep{Gaume}. There are also two young massive clusters (YMCs) -- Arches and Quintuplet -- which have formed in the CMZ recently, with ages of \ap 3.5 and 4.8 Myr, respectively \citep{Schneider14}. These clusters contain \ap 10$^{4}$~M$_{\odot}$ of stars, making them amongst the most massive clusters to have recently formed in the Galaxy \citep{ymc_port}.

Searching for the precursors to clusters like the Arches and Quintuplet, which contain tens of thousands of stars, allows us to study how these stars form in this extreme environment. Sagittarius B2 is a likely candidate for such a precursor system. However, the intense ionising radiation from the forming high-mass stars has disrupted the initial gas conditions. If we are to understand the early stages of star formation at the Galactic centre, we need to find gas clouds in which the initial structure has yet to be perturbed by widespread star formation.

Given a star formation efficiency of $\lesssim$~30\%, we expect that the precursors to clusters like the Arches and Quintuplet should be molecular clouds of order 10$^{5}$~M$_{\odot}$, contained within several parsecs, such that a high-mass (10$^{4}$~M$_{\odot}$) and compact (R$_{core}$ \ap 0.1~pc) cluster can form. A growing sample of such clouds have been identified throughout the Galaxy \citep[see e.g.][]{Bricklets, ymc_steve_14, Ginsburg_clouds, Ginsburg_w51_16, atlasgal_hii}. All of the YMC precursor candidates that have been identified in the Galactic disc are seen to be forming stars at a high rate (e.g. W49, W51). This is in agreement with the threshold proposed by \citet{Lada10}. The YMC precursor clouds that have been identified at the Galactic centre, however, are mostly quiescent -- devoid of any widespread star formation (the notable exception being the highly-star-forming Sagittarius B2 complex), despite their masses, sizes and densities being similar to those found in the disc. This difference in star-forming activity highlights that this single volume density threshold for star formation may not hold in the more extreme environment of the Galactic centre, which has now also been found to be true in the centres of other galaxies \citep{usero15,bigiel16}.

Many of the known quiescent YMC precursor candidates near the Galactic centre are confined to a relatively small region, which is situated in a gas stream spanning \ap 0.3 degrees (\ap 45~pc at a distance of 8.4~kpc) in projection. This region, known as the \q{dust ridge} \citep{Bricklets_lis}, contains six prominent infrared dark clouds. These are G0.253$+$0.016 (aka \q{the Brick}), G0.340$+$0.055, G0.380$+$0.050, G0.412$+$0.052, G0.478$-$0.005 and G0.496$+$0.020. For the sake of brevity, we adopt the nomenclature of \citet{Bricklets_lis} and will hereafter refer to these clouds as \q{a -- f}, respectively. Of these, four clouds, \q{a}, \q{d}, \q{e} and \q{f} have been identified as potential progenitors to YMCs \citep{Bricklets, rathborne15, Walker15}. Clouds \q{b} and \q{c}, whilst not massive enough to be considered YMC precursors, still contain tens of thousands of solar masses of gas and dust within only a few parsecs \citep{Immer} and therefore still have the potential to form substantial star clusters.

Given that we expect to find a significant number of high-mass stars in massive stellar clusters\footnote{A greater number of high-mass stars are expected in massive stellar clusters as a result of greater sampling of the stellar initial mass function.}, these dust ridge clouds provide an excellent laboratory in which to probe the early, largely unperturbed conditions of both massive star and cluster formation in the extreme environment of the CMZ. 

In this paper, we present high angular resolution observations of these clouds from the Submillimeter Array Legacy Survey of the Central Molecular Zone \citep[{\it{CMZoom}},][]{cmzoom_cara}, the details of which are discussed in section \hyperref[sec:Data]{2}, along with a discussion of the complimentary single-dish data that we use to recover the large scale emission that is filtered out by the interferometer. In section \hyperref[sec:Results]{3} we present the analysis of the dust continuum and a select few molecular lines, which we use to determine the number of dense cores embedded within the clouds along with their general properties such as masses, radii, gas temperatures, line-widths and virial ratios. The results of this analysis are presented in Table \hyperref[table:Table2]{3}. In section \hyperref[sec:Discussion]{4} we compare the properties of the dense cores that we have identified in the CMZ with those of cores in the less extreme environment of the Galactic disc to investigate any potential effects these differing environmental conditions may have on the embedded core populations of molecular clouds. The conclusions and implications of this study are presented in sections \hyperref[sec:Discussion]{4} and \hyperref[sec:Conclusion]{5}. The analysis of the molecular line data and methodology of the gas temperature modelling are given in appendix \hyperref[A.1]{A}.

\section{Observations \& Data} \label{sec:Data}
\subsection{\emph{CMZoom}\protect\footnote{https://www.cfa.harvard.edu/sma/LargeScale/CMZ/} -- An SMA Legacy Survey of the Central Molecular Zone}\

In an effort to understand the physical, chemical and kinematic properties of the dense structure in the CMZ, we have embarked on a large legacy survey \citep[{\it{CMZoom}},][]{cmzoom_cara} of the region with the Submillimeter Array (SMA). We have completed \ap 500 hours worth of mapping, corresponding to \ap 240~arcmin$^{2}$ of the CMZ at 230 GHz (1.3~mm). The survey was designed to target all regions within the CMZ that lie above a column density threshold of $\gtrsim$~10$^{23}$~cm$^{-2}$. The typical spatial resolution of this survey is \ap 4" (0.16~pc at a distance of 8.4~kpc) and the spectral resolution is \ap 1.1~km s$^{-1}$.

The dust ridge clouds from \q{b -- f} (see Table 1 for cloud properties) were observed in the compact array configuration between the 24\ts{th} May -- 6\ts{th} June 2014, and in the subcompact configuration between the 25\ts{th} -- 27\ts{th} July 2014 (see Table 2 for more details of the observations). Cloud \q{a} was not observed, as it is well-studied and high-resolution data already exist \citep{Brick_jens, Brick_KJ, Jill_pdf_2014, rathborne15}. Typical RMS continuum sensitivity achieved is \ap 3 -- 5 mJy beam$^{-1}$. Figure 1 shows a 3-colour image of the dust ridge, where white circles correspond to the primary beam coverage of the SMA observations.

Data calibration was performed using {\sc mir}\footnote{{\sc mir} is an IDL-based package that has been developed to calibrate SMA data. The {\sc mir} package and cookbook can be found at https://www.cfa.harvard.edu/$\sim$cqi/mircook.html.}, continuum and line separation was done using {\sc miriad} \citep{miriad} and all subsequent imaging and analysis was performed using {\sc casa} \citep{casa}.

%___________
\begin{table}
\begin{center}
  \label{tab:global_properties}
  \begin{tabular}{ccccc}
    \hline
    Cloud	&	Mass 	&	Radius & $n$ & Reference\\
    & 	10$^{4}$ M$_\odot$	&	pc & 10$^{4}$ cm$^{-3}$ & \\ \hline
    b & 1.3 & 1.9 & 0.7 & 2\\
    c & 1.8 & 1.9 & 0.9 & 2\\ 
    d & 7.6, 7.2 & 3.2, 3.4 & 0.8 & 1, 2\\ 
    e & 11.2, 15.3 & 2.4, 4.5 & 2.8 & 1, 2\\  
    f & 7.3, 7.2 & 2.0, 2.7 & 3.2 & 1, 2\\ 
    \hline \hline
  \end{tabular}
  \caption{Masses, radii and mean number densities (assuming spherical geometry) of the clouds that we have observed with the SMA as part of the larger survey. The relevant references for these quantities are also given. References: (1) \citet{Walker15} and (2) \citet{Immer}.}
\end{center}
\end{table}
%___________

Although we have both compact and subcompact SMA observations, the data still suffer from the inherent limited flux recovery due to the incomplete {\it{uv}}-coverage of the interferometer. To account for this, we have obtained the most appropriate single-dish data for combination with the interferometric data.

To recover the continuum emission from these clouds, we use data from the Bolocam Galactic Plane Survey \citep[BGPS v2.1,][]{bgps1,bgps2,bgps3}. These data are at a wavelength of 1.1~mm and provide a beam resolution of 33'' (\ap 1.3~pc at a distance of 8.4~kpc) and resolve a largest angular scale of \ap 120$^{\prime\prime}$ \citep{bgps3}. Ideally, we would use single-dish observations that were taken at 1.3~mm -- the wavelength of our SMA data. Given that we have only single-dish data at 1.1~mm, we scale the data to estimate the flux that would be observed at 1.3~mm, which is determined by taking the ratio of the fluxes ($S_{\nu}$) at each wavelength, with an assumed emissivity index of $\beta$ = 1.75. 

We use the \emph{feather} task in {\sc casa} when combining the BGPS data with the cleaned SMA data. Any discussion regarding the dust continuum hereafter refers to results obtained using fully combined maps of compact, subcompact and single-dish data.

For the molecular line data, we use data from a recent survey of the CMZ performed using the Atacama Pathfinder EXperiment (APEX) telescope \citep{Adam_cmz_temp}. The method\footnote{http://tinyurl.com/zero-spacing} used to combine the single-dish and interferometer line data are somewhat more involved than for the case of the continuum. We first have to ensure that the APEX data are formatted correctly and that they are gridded to the same axes as the SMA data. A first iteration of {\sc clean} is performed with the APEX data as a model, the output of which is then used to create an updated model image. This updated model image is then used in a final iteration of {\sc clean} to produce the combined data cube.

\begin{figure*}
\begin{center}
\includegraphics[scale=0.38, angle=-90]{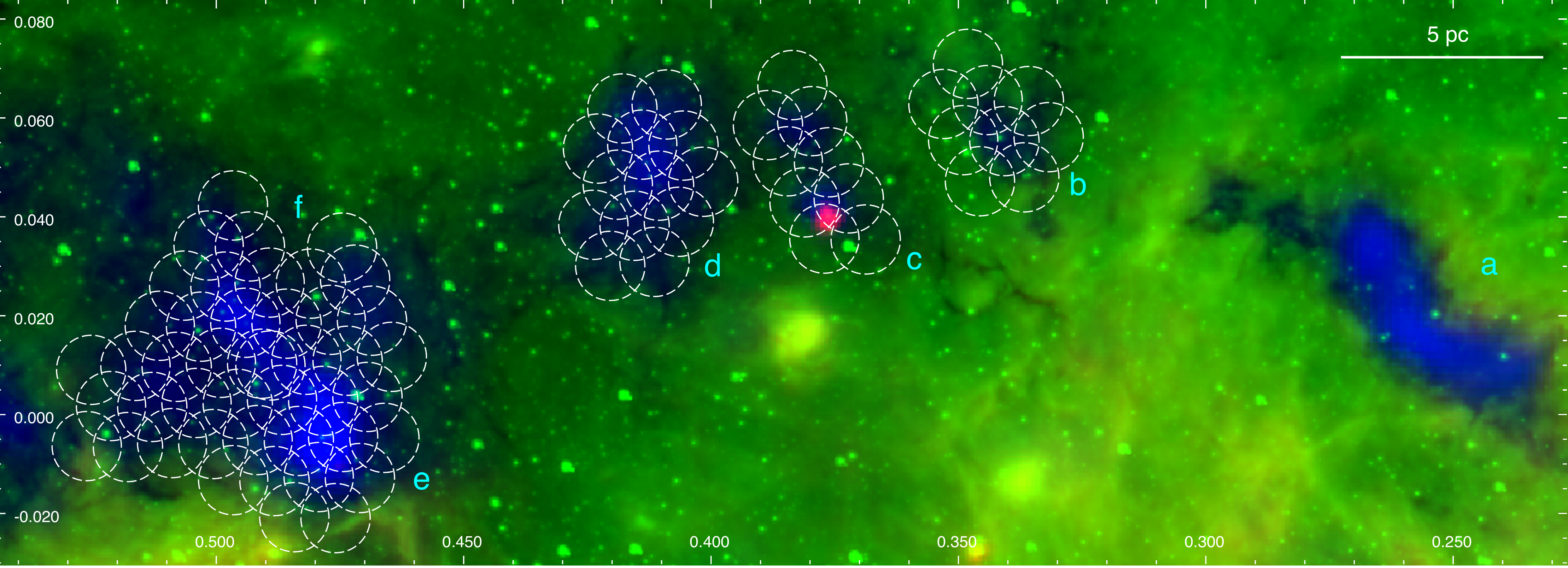}
 \caption{Three-colour image of the \q{dust ridge} at the Galactic centre. Green: 8 $\mu$m data from the Spitzer/GLIMPSE survey \citep{glimpse}. Blue: the column density map from the HiGAL survey \citep[][range displayed is \ap 5 -- 55 x 10$^{22}$ cm$^{-2}$]{higal, Cara_higal}. Red: Herschel 70 $\mu$m emission. White circles correspond to the primary beam coverage of our SMA observations.}
\end{center}
\end{figure*}

%___________
\begin{table*}
\begin{center}
  \label{tab:global_properties}
  \begin{tabular}{ccccccc}
    \hline
    Cloud & Central coordinates	& Resolution &	N$_{pointings}$ 	&	N$_{antennas}$ & Time on source & RMS\\
    & & ($^{\prime\prime}$) & &	& (hours) & (mJy beam$^{-1}$)\\ \hline
    b & G0.340+0.055 & 3.18 $\times$ 3.13 & 9 & 7 & 2.36 & 2--3\\
    c & G0.380+0.050 & 3.15 $\times$ 3.10 & 9 & 7 & 2.30 & 2--5\\ 
    d & G0.412+0.052 & 3.19 $\times$ 3.08 & 13 & 7 & 3.04 & 3--5\\ 
    e/f & G0.489+0.010 & 3.04 $\times$ 2.93 & 44 & 7 & 11.35 & 2--6\\  
    \hline 
    \hline
  \end{tabular}
  \caption{Details of our SMA observations of dust ridge clouds b -- f. All clouds were observed using the ASIC correlator with a total bandwidth of 8~GHz, between 217.05 -- 232.83~GHz. Spectral resolution is 1.1~km~s$^{-1}$ in all cases.}
\end{center}
\end{table*}
%___________

\section{Results} \label{sec:Results}
\subsection{1.3mm Continuum Emission}\

Figure 2 displays the individual 1.3~mm continuum maps for each of the clouds, overlaid as black contours on top of Herschel column density maps (blue). We clearly detect dense substructure in each of these clouds. In order to describe this substructure in a systematic way, we produce dendrograms \citep[see e.g.][]{dendro} for each of the clouds using the {\sc astrodendro}\footnote{{\sc astrodendro} is a Python package designed to compute dendrograms of astronomical data and can be found at http://www.dendrograms.org/.} software package. To do this, we set a threshold of 5$\sigma$ and an increment of at least 2$\sigma$ between structures (see Table 2 for RMS estimates). We specify the minimum number of pixels required as --
\smallskip
\begin{equation}
N_{pix}(min) = \frac{2 \pi \theta_{maj} \theta_{min}}{8ln(2)A_{pix}}
\end{equation}
\smallskip

\noindent where $\theta_{maj}$ and $\theta_{min}$ are the major and minor axes of the synthesised beam and $A_{pix}$ is the pixel area. For example, for respective values of 3.2$^{\prime\prime}$, 3.1$^{\prime\prime}$ and 1$^{\prime\prime^{2}}$, we obtain $N_{pix}(min)$ \ap 11 pixels for the observation of dust ridge cloud \q{c}.

We detect a total of 15 independent cores throughout clouds \q{b -- f}. These are shown as red contours in Figure 2. These cores, along with their estimated fluxes, effective radii (R$_{eff}$), positions and masses are given in Table 3. {\sc astrodendro} calculates the radius by taking the geometric mean of the major and minor axes of the projection onto the position-position plane, computed from the intensity-weighted second moment in the direction of elongation. However, the ellipses that are fitted to the cores often appear to underestimate the radius. We instead use the total area of each core and calculate the effective radius of a circular source with an area equal to that of the core. The difference between these radii can be as large as 30 -- 40\%. Our radius estimates are therefore upper limits.

In order to estimate the masses of the detected cores, we use the following relation \citep{Jens_mass_in} --

\begin{equation}
M = \frac{d^2}{\kappa_{\nu} B_{\nu}(T)} \int I_{\nu} d\Omega = \frac{d^2 F_{\nu}}{\kappa_{\nu} B_{\nu}(T)}
\end{equation}
\smallskip

\noindent which can be written in the more-readily useable form -- 

\begin{equation}
\begin{split}
M = 0.12 M_{\odot}  \left( {\rm e}^{1.439 (\lambda / {\rm mm})^{-1}
      (T / {\rm 10 ~ K})^{-1}} - 1 \right) \\
   \times \left( \frac{\kappa_{\nu}}{0.01 \rm ~ cm^2 ~ g^{-1}} \right)^{-1} 
   \left( \frac{F_{\nu}}{\rm Jy} \right)
  \left( \frac{d}{\rm 100 ~ pc} \right)^2
  \left( \frac{\lambda}{\rm mm} \right)^{3}
\end{split}
\end{equation}

\bigskip

\noindent where $M$ is the mass, $\lambda$ is the wavelength, $T$ is the dust temperature, $\kappa_{\nu}$ is the dust opacity, $F_{\nu}$ is the integrated flux and $d$ is the distance. The distance to all sources is taken to be 8.4~kpc \citep{Distance, Reid_2014}. The dust opacity ($\kappa_{\nu}$) is not observationally constrained towards these cores, and so we instead estimate this using the following relation  \citep{Cara_higal} -- 

\begin{equation}
\kappa_{\nu} = 0.04 \rm ~ cm^{2} ~ g^{-1} \left(\frac{\nu}{505 ~ GHz}\right)^{1.75}
\end{equation}

\bigskip

\noindent where $\nu$ is the frequency. Note that this contains the explicit assumption that the gas-to-dust ratio is 100, which may not be valid in the CMZ \citep{snl_sf}. \citet{Jens_mass_in} note that the uncertainties in both the dust temperature and opacity mean that the systematic uncertainties in mass estimates obtained via Equation 3 are a factor of \ap 2. See \citet{snl_sf} for a more in-depth discussion regarding the systematic uncertainties in obtaining mass estimates from dust emission in this environment.

%___________________________________________________________________
    \begin{figure*}
        \centering
        \begin{subfigure}[b]{0.475\textwidth}
            \centering
            \includegraphics[width=\textwidth, angle=-90, scale=0.85]{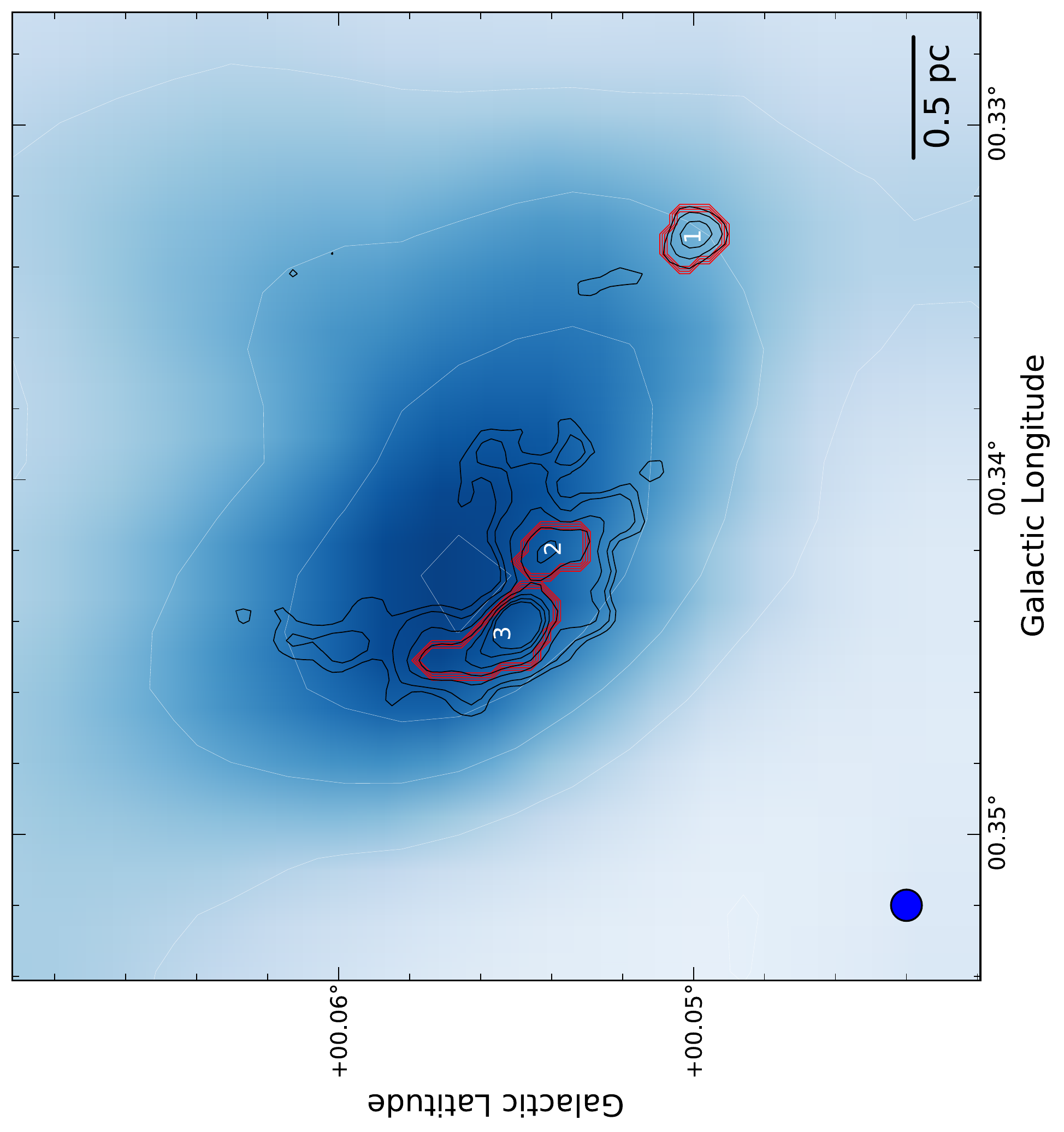}
            \caption[]%
            {{\small}}    
        \end{subfigure}
        \hfill
        \begin{subfigure}[b]{0.475\textwidth}  
            \centering 
            \includegraphics[width=\textwidth, angle=-90, scale=0.85]{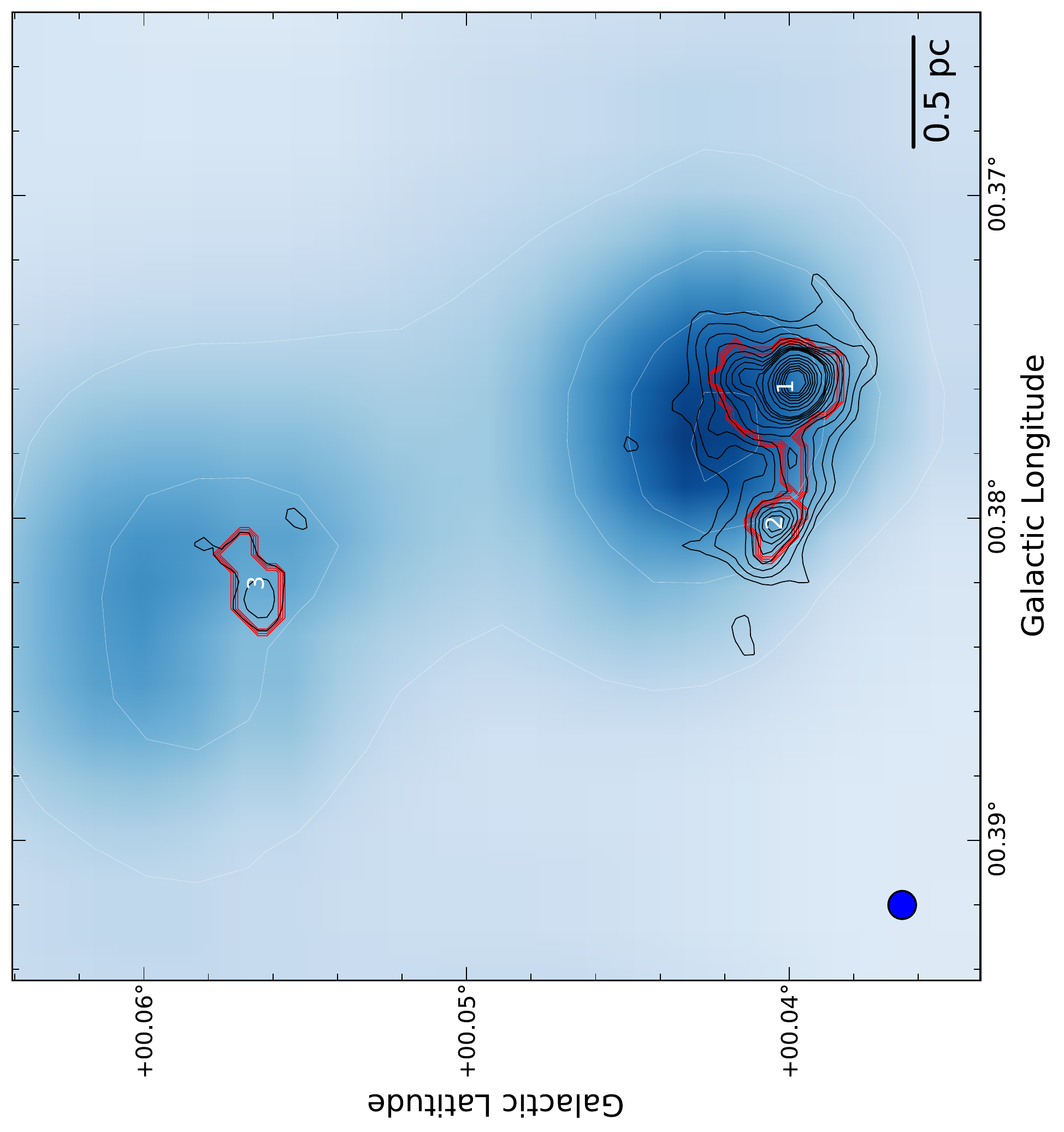}
            \caption[]%
            {{\small}}    
        \end{subfigure}
        \vskip\baselineskip
        \begin{subfigure}[b]{0.475\textwidth}   
            \centering 
            \includegraphics[width=\textwidth, angle=-90, scale=0.85]{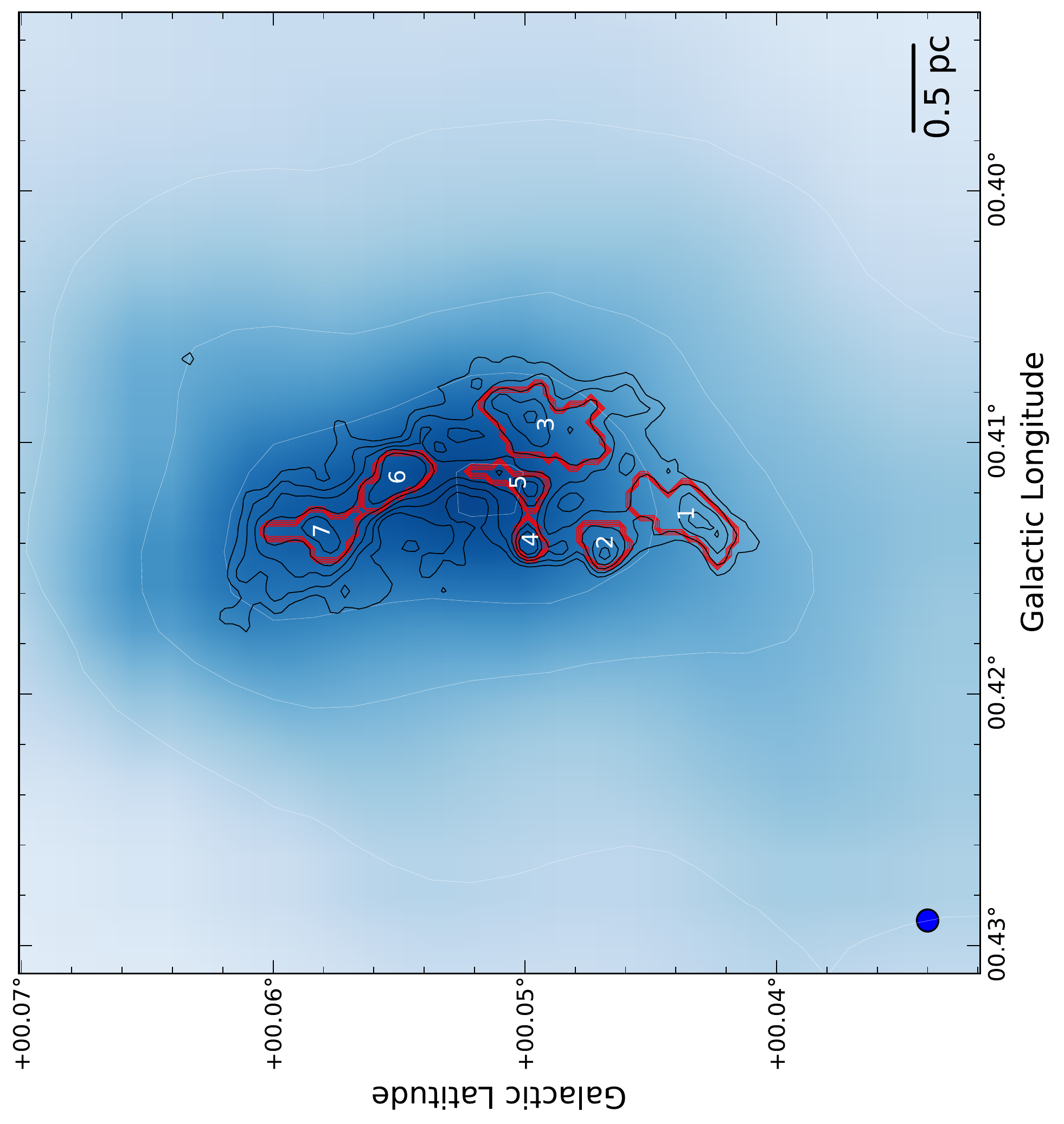}
            \caption[]%
            {{\small}}    
        \end{subfigure}
         \quad
        \begin{subfigure}[b]{0.475\textwidth}   
            \centering 
            \includegraphics[width=\textwidth, angle=-90, scale=0.85]{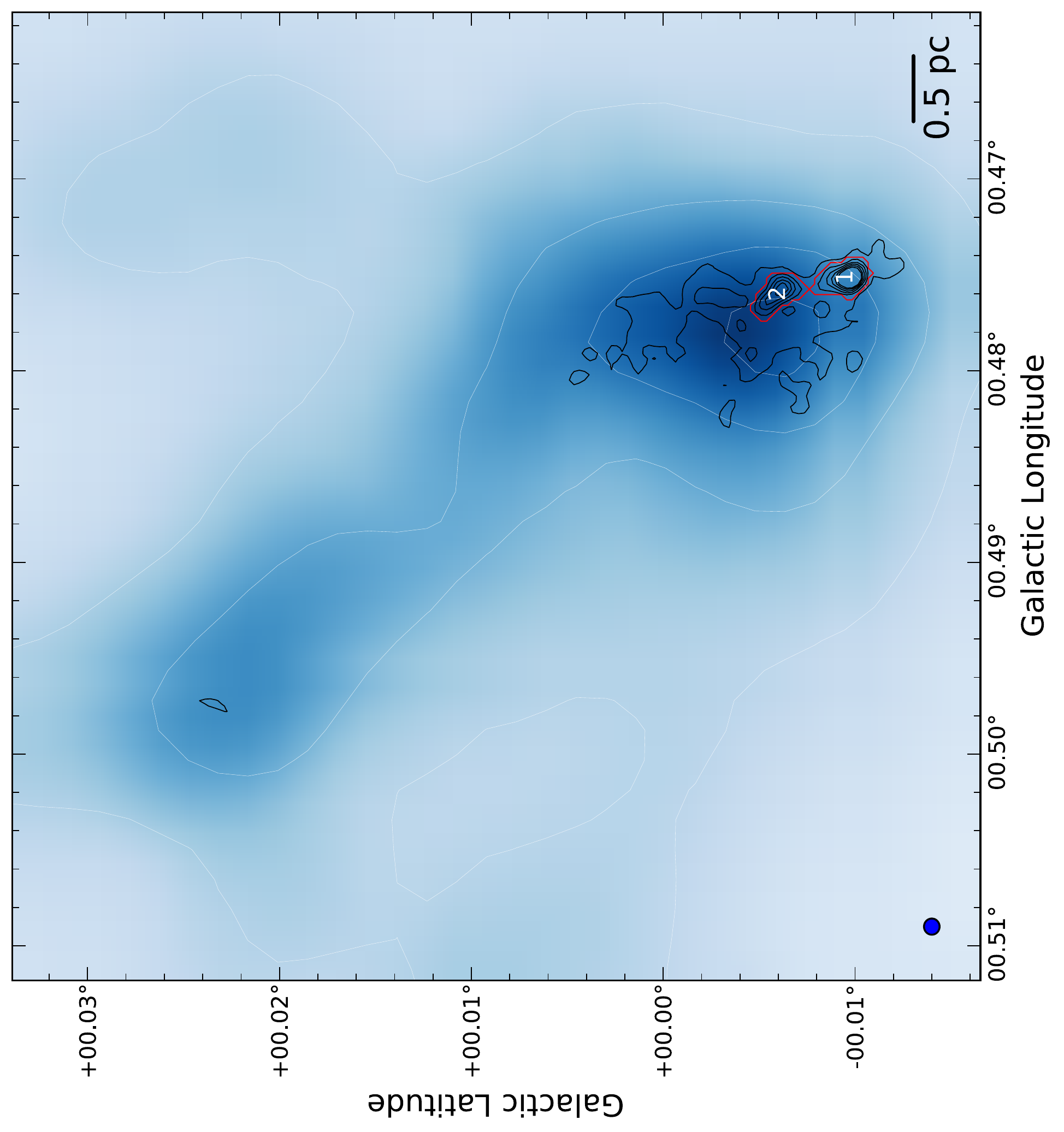}
            \caption[]%
            {{\small}}    
        \end{subfigure}
        \caption[]
        {\emph{Top:} dust ridge clouds \q{b} (\emph{left}) and \q{c} (\emph{right}). \emph{Bottom:} dust ridge clouds \q{d} (\emph{left}) and \q{e/f} (\emph{right}). Background image is a Herschel column density map, shown with corresponding white contours. Black contours show the 230~GHz continuum emission as seen with the SMA at the 5-$\sigma$ level. They are 10 -- 32, , 16 -- 32, 16 -- 460 and 20 -- 72 mJy beam$^{-1}$ for clouds b, c, d and e/f, respectively. The synthesised beam is shown in the lower left of each panel. Red contours highlight the cores as determined via dendrogram analysis.} 
    \end{figure*}
%___________________________________________________________________

Besides the uncertainty of the gas-to-dust ratio in the CMZ, the only other parameter in Equation 3 that remains unconstrained is the dust temperature towards these cores. At present, we do not know the temperature of the dust on the scales that we are probing with our SMA observations (\ap 4"). The best measure of the dust temperature that we do possess is from the HiGAL survey \citep[][temperature estimates by \citealt{Cara_higal}]{higal}. This provides us with dust temperatures that are typically \ap 20~K in these clouds on 33" scales. We therefore use this dust temperature to provide an upper limit to the masses of these cores. This is an upper mass limit due to the fact that a decrease in temperature results in an increase in the measured mass, as can be seen in Equation 3. We expect that the gas and dust may be internally heated towards some of these cores, which would consequently drive the mass estimates down. Though it is also possible that a quiescent embedded core with no internal heating may be even colder than the gas on larger scales due to being heavily shielded from external heating sources.

We find, assuming this dust temperature of 20~K, that the cores we have detected with the SMA have upper mass limits ranging from \ap 50 -- 2150~M$_{\odot}$ within radii of \ap 0.1 -- 0.25~pc (see Table 3) and have volume densities \ap 10$^{6}$~cm$^{-3}$. As previously discussed, these are likely upper mass limits. In the next section we discuss how we use molecular line data to determine gas temperatures and infer lower mass limits for the cores. 

\begin{table*}
	\tabcolsep=0.16cm
	\begin{tabular}{ccccccccccc}
		\hline
		Source & Enclosed Flux & R$_{eff}$ & $\Delta l$ & $\Delta b$ & M$_{T_{20}}$ & $<$T$_{gas}$$>$ & M$_{T_{gas}}$ & $\Delta$V & $\alpha$$_{vir}$\\
		& (Jy) & (pc) & ($^{\prime\prime}$) & ($^{\prime\prime}$) & (M$_{\odot}$) & (K) & (M$_{\odot}$) & (km s$^{-1}$) & \\ \hline 
		G0.340+0.055\\ \hline
b1 &		0.03 & 0.12 & 23.53 & -18.92 & 51 & -- & -- & -- & --\\
b2 &		0.05 & 0.12 & -8.15 & -4.80 & 78 & -- & -- & -- & --\\
b3 &		0.17 & 0.19 & -16.75 & 0.46 & 251 & -- & -- & -- & --\\ \hline

		G0.380+0.050\\ \hline
c1 &		1.45 & 0.26 & 13.56 & -36.73 & 2143 & {\textcolor{black}{220.3}} & {\textcolor{black}{140}} & 5.1 $\pm$ 0.2 & 0.7 -- {\textcolor{black}{10.1}} \\
c2 &		0.16 & 0.12 & -1.95 & -35.12 & 236 & \textless \ 56.7 & \textgreater \ 68 & 5.4 $\pm$ 0.6 -- 6.0 $\pm$ 0.5 & 3.1 -- 13.3 \\
c3 &		0.09 & 0.15 & -7.92 & 22.97 & 135 & -- & -- & -- & -- \\ \hline

		G0.412+0.052\\ \hline
d1 &		0.19 & 0.22 & -5.24 & -32.10 & 285 & -- & -- & -- & --\\
d2 &		0.08 & 0.13 & -8.67 & -19.62 & 119 & 86.3 & 22 & 4.2 $\pm$ 0.3 -- 5.0 $\pm$ 0.3 & 4.0 -- 30.9\\
d3 &		0.31 & 0.26 & 8.51 & -11.39 & 455 & -- & -- & -- & -- \\
d4 &		0.05 & 0.09 & -8.03 & -8.94 & 71 & -- & -- & -- & --\\
d5 &		0.07 & 0.12 & -0.70 & -9.05 & 109 & -- & -- & -- & --\\
d6 &		0.16 & 0.16 & 1.05 & 10.16 & 239 & \textless \ 56.7 & \textgreater \ 69 & 5.4 $\pm$ 0.5 -- 6.3 $\pm$ 0.4 & 4.1 -- 19.0\\
d7 &		0.15 & 0.17 & -6.74 & 21.30 & 225 & -- & -- & -- & --\\ \hline

		G0.489+0.010\\ \hline
e1 &		0.30 & 0.16 & 48.55 & -71.03 & 445 & $>$ 150 & $<$ 46 & 5.4 $\pm$ 0.6 -- 7.8 $\pm$ 0.6 & 2.1 -- 44.1\\
e2 &		0.22 & 0.15 & 45.57 & -58.27 & 325 & -- & -- & -- & -- \\ \hline
	\end{tabular}
	\caption{The above table displays the dust cores that have been identified using the dendrogram analysis. For each cloud, the table lists the cores identified, along with the enclosed flux, effective radius (R$_{eff}$), central co-ordinates as offsets ($\Delta l$, $\Delta b$, with respect to central coordinates given in the first column) and the core mass assuming a dust temperature of 20~K (M$_{T_{20}}$). Where possible, we also provide the spatially-averaged gas temperature (see \hyperref[sec:Line]{3.2}), core mass assuming a dust temperature equal to the gas temperature (M$_{T_{gas}}$), line-width of H$_{2}$CO ($\Delta$V) and the virial ratio ($\alpha$$_{vir}$). We provide two measures of the line-width -- one with and one without the inclusion of single-dish data. All calculations assume a gas-to-dust ratio of 100 and a distance of 8.4~kpc \citep{Distance, Reid_2014}.}
\label{table:Table2}
\end{table*}

\subsection{Molecular Line Emission}\ \label{sec:Line}

Currently, we only have a lower limit on the dust temperature from Herschel observations (\ap 20~K on 33$^{\prime\prime}$ scales). It is known that on these larger scales, the gas and dust temperatures in the CMZ are not thermalised and that the gas temperatures are typically higher \citep[see e.g.][]{Ao_gas_temp, Brick_KJ, Adam_cmz_temp, Immer16}. However, it is possible that the gas and dust may begin to thermalise at high volume densities. \citet{Clark_brick} show this in Figure 3 of their paper, in which they model the temperature of the gas and dust as a function of volume density, showing that they begin to couple at densities \textgreater \ 10$^{6}$~cm$^{-3}$ and converge at \textgreater \ 10$^{7}$~cm$^{-3}$. Their models assume cosmic ray ionisation rates and interstellar radiation fields in the CMZ that are much higher than those measured in the local interstellar medium. For more typical, local conditions, the gas and dust would be expected to couple at lower densities, \ap 10$^{4}$~cm$^{-3}$ \citep[e.g.][]{Goldsmith_temp, Galli_temp}.

The cores that we present here are at volume densities of \ap 10$^{5}$ -- 10$^{6}$~cm$^{-3}$, based on their upper-limit mass estimates. We therefore reason that if we can estimate the temperature of the \emph{gas}, this can be used to provide strong upper limits for the \emph{dust} temperature and hence lower limits for the core masses. However, we note that based upon the models of \citet{Clark_brick}, there may still be a difference between the gas and dust temperatures at these densities of several factors. Thus, any mass estimates made this way are likely extreme lower limits.

For the purposes of this paper, we utilise the following molecular lines that were covered in our SMA bands -- the triplet of lines from the para-H$_{2}$CO (Formaldehyde) transitions at 218.22219 (3$_{0,3}$-2$_{0,2}$), 218.47563 (3$_{2,2}$-2$_{2,1}$) and 218.76007 (3$_{2,1}$-2$_{2,0}$)~GHz as well as the J=12-11 k-ladder of CH$_{3}$CN (Methyl Cyanide). We choose to use these transitions as their emission lines can be used as a thermometer to estimate the temperature of the gas from which they are emitted (see e.g. \citealt{Ao_gas_temp, Brick_KJ, Adam_cmz_temp, Immer16} for H$_{2}$CO temperature measurements in the CMZ and \citealt{Longmore_ch3cn} for CH$_{3}$CN temperature measurements towards a high-mass star forming region). A more detailed description of how we utilise these transitions for estimating temperatures is presented in the appendix of this paper.

Of the 15 cores detected with the SMA, we are able to place constraints on the gas temperatures for five of them. The remainder either had poor signal-to-noise or non-detection. In {\textcolor{black}{Table 3}}, we present the estimated gas temperature ($<$T$_{gas}$$>$), which has been averaged over the spatial extent of each core, as specified by their measured radii. We then assume this as the average dust temperature and re-calculate the mass of each core in the same way as in the previous section (M$_{T_{gas}}$). We find a varying degree of gas temperatures in cores, from \ap 57~K up to \ap {\textcolor{black}{220~K}}. Assuming equivalent dust temperatures has a substantial effect on the mass estimates, decreasing by as much as {\textcolor{black}{93\%}} in the most extreme case.

Though this provides strong lower-mass estimates for the cores, the significant variation in estimated mass highlights the need for accurate estimates of the dust temperature on these small spatial scales. The de-coupled nature of the gas and dust temperatures in the CMZ makes this difficult observationally. Radiative transfer modelling of the dust and line emission is required such that we can determine the volume densities of these cores. Knowing the density, we can then compare this with numerical models of the thermal coupling of the gas and dust in the CMZ as a function of volume density (see e.g. Figure 3 of \citealt{Clark_brick}). However, we do not possess the necessary line data to do this currently. We find that the ratio of the (3$_{2,1}$-2$_{2,0}$)/(3$_{0,3}$-2$_{0,2}$) lines of H$_{2}$CO, which we use as a primary temperature diagnostic, often exceeds that which can be reliably fit by any models at high volume density (see appendix). We are therefore only able to determine lower limits in such cases.

We also use the line-widths from our line-fitting to investigate whether the cores are likely to be gravitationally bound. To do this we calculate the virial parameter ($\alpha$ = $5R\sigma^2$/$GM$, where $\sigma$ = $\Delta$V/2$\sqrt{2\textrm{ln}(2)}$, $\Delta$V is the line-width and $\alpha$ $\lesssim$ 2 typically indicates gravitationally bound) for both the upper and lower mass estimates. We also use two values for the line-width ($\Delta$V). These two values come from fitting the H$_{2}$CO lines with and without single-dish data added. The single-dish APEX data are sensitive to the larger scale emission in these clouds, which is known to display very broad line-widths (10 -- 20~km~s$^{-1}$). This means that in general, the single-dish combination tends to broaden the lines. For the purposes of estimating whether or not the cores may be gravitationally bound, it is not clear that including the single-dish data is appropriate here. We therefore present both values for $\Delta$V and use these to estimate upper and lower limits for $\alpha$.

Of the five cores for which we have H$_{2}$CO detection, we find that only two of them, \q{c1} and \q{e1} , have $\alpha$ $\lesssim$ 2 based on their upper mass limits. The other three cores have $\alpha$ \ap 3 -- 4, suggesting that they may be unbound. Re-calculating $\alpha$ with the lower mass and larger line-width estimates puts it at much higher values, up to as large as 44, suggesting that they are unbound. However, we again caution that these are extreme lower limits for the mass and upper limits for the line-width. This further reinforces the need for more accurate dust temperatures on the scale of these cores. {\textcolor{black}{We also note that using this virial parameter estimation as a measure of boundedness is likely too much of an oversimplification in the case of these sources, where it is possible that other physical mechanisms may play a significant role. We explore this in more detail in $\S$4.2, with a particular focus on the significance of external pressure in the extreme environment of the CMZ.}}

{\textcolor{black}{In summary, while the cores that we have detected are massive and compact, they do not appear to be strongly sub-virial as a result of their large line-widths of \ap 4 -- 6~km~s$^{-1}$. Naively, one might assume that such broad line-widths on these small scales is due to the heightened turbulence in the CMZ, as is thought to be the case for the very broad (10 -- 20~km~s$^{-1}$) line-widths on large spatial scales in this region \citep{Henshaw_cmz}. However, it is not clear that the large-scale turbulence in the CMZ should significantly broaden the line-widths on these core-scales, and it is also not possible to exclude other options (i.e. infalling/outflowing material) without a more detailed analysis of the kinematics. Additionally, \citet{Jens_16a, Jens_16b} recently presented a survey of molecular clouds in the CMZ, and do indeed find evidence of narrow (\textless \ 2~km~s$^{-1}$) velocity dispersions on small scales in the CMZ. They conclude that objects on the scale of the cores presented in this paper (\ap 0.1~pc) in their sample have typical virial ratios of $\lesssim$ 2 -- 3, and should therefore be close to being gravitationally bound. This is broadly consistent with the lower values presented here, based upon upper mass limits. However, the line-widths that we calculate are typically larger than those in the \citet{Jens_16a, Jens_16b} sample, though this may be a result of our choice of molecular line tracer, as they use N$_{2}$H$^{+}$, whereas we are using H$_{2}$CO. Further studies of the line emission from all of these cores, as well as those discovered in the wider \emph{CMZoom} survey, will provide a more detailed insight as to whether this difference is a real one, or whether it is dependent upon the choice of tracer.}}

\bigskip

\section{Discussion} \label{sec:Discussion}

\subsection{Young precursors to high-mass stars?}\

As explained in the introduction ($\S$1), these dust ridge clouds are intriguing in that their density lies at or above the threshold of density-dependent star formation relations (\ap 10$^{4}$~cm$^{-3}$, \citealt{Lada10}), yet they are devoid of widespread star formation \citep{snl_sf}. Clouds with similar global properties in the less extreme environment of the Galactic disc are all forming stars prodigiously \citep[see e.g.][]{Ginsburg_clouds, atlasgal_hii}. This raises the question -- if star formation is being inhibited at these densities in the CMZ and pushing the critical density for star formation up to higher values, will these clouds ever form stars/clusters? Indeed, this has already been a subject of debate regarding G0.253$+$0.016 \citep[cloud \q{a}, see e.g.][]{Brick, Brick_jens, Brick_KJ, Jill_pdf_2014}.

Whether or not these clouds will form stellar clusters is still uncertain, but we know that star clusters have recently formed in the CMZ (e.g. the Arches and Quintuplet) and that clusters are currently forming there (e.g. Sagittarius B2). It is also predicted that the environmental conditions in the CMZ should lead to a higher fraction of stars forming in bound clusters \citep{Diederik_cfe}. These (proto-)clusters must have an earlier, quiescent phase, prior to the onset of widespread star formation. To date, these dust ridge clouds are some of the best candidates for such a phase in the CMZ. If these clouds are precursors to such clusters, we expect that they will contain the precursors to high-mass stars. Our SMA observations reveal 15 dense, high-mass cores within these clouds that may be the potential formation sites of high-mass stars and sub-clusters. 

{\textcolor{black}{Of these 15 detected cores, 13 do not show any clear signs of ongoing star formation. This is based upon a search for H$_{2}$O and CH$_{3}$OH Class II maser emission, along with a search for 24~$\mu$m emission \citep[Spitzer, GLIMPSE,][]{glimpse} and 70~$\mu$m emission \citep[Herschel, HiGAL,][]{higal}. We also do not see any significant molecular line emission towards these 13 cores in the SMA sidebands, and detected no obvious outflow signatures, though we did not explicitly search for them.}} Detailed analysis of the molecular line data is required to determine the fate of these cores. Based on our analysis here, we see that the properties of the star-forming and the quiescent cores derived from the SMA continuum and line data are similar in terms of their masses, sizes, densities and line-widths, suggesting that they all have the capacity to form stars. In this case, they would represent different evolutionary phases of the same type of object. Our estimated ranges of the virial ratio (see Table 3) would suggest that some of the cores are unbound and will therefore not form stars. However, these are highly uncertain due to the lack of constraints on the dust temperature. It is also possible that some of the cores may be unbound on the scales probed by our SMA observations, but that smaller scale structure (i.e. on which individual stars form) within the cores could be gravitationally bound.

The two other cores -- \q{c1} and \q{e1} -- show signs of star formation activity. \q{e1}, situated towards the south of cloud \q{e}, has the second highest peak brightness of all 15 cores (\ap 0.12 Jy beam$^{-1}$) and has a mass in the range of \ap 46 -- 445~M$_{\odot}$ within a radius of 0.16~pc. This source coincides with weak 70~$\mu$m emission as seen with Herschel and both H$_{2}$O and CH$_{3}$OH Class II maser emission \citep{c_maser}. 

\q{c1}, situated towards the south of cloud \q{c}, has the highest peak brightness in the sample (\ap 0.6 Jy beam$^{-1}$) and has a mass in the range of \ap {\textcolor{black}{140}} -- 2143~M$_{\odot}$ within a radius of 0.26~pc. We find that this source coincides with strong 70~$\mu$m emission as seen with Herschel and both H$_{2}$O and CH$_{3}$OH Class II maser emission \citep{c_maser}. Additionally, this core coincides with H$_{2}$CO and SiO maser emission -- both of which are extremely rare in star-forming regions in the Galaxy \citep{Adam_c}. Thus far, only eight H$_{2}$CO masers have been identified in the Galaxy, all of which are associated with high-mass star formation. Even more rare are SiO masers. They are commonly detected towards evolved stars, but only rarely towards star-forming regions, and then only in regions of known high-mass star formation \citep{SiO_masers}. Thus far, only five SiO masers have been identified towards such regions in the Galaxy. Furthermore, H$_{2}$CO and SiO masers have only been simultaneously detected towards two regions -- Sagittarius B2 and \q{c1} -- both of which are in the CMZ. This may also suggest something different about the star formation and/or chemistry in the CMZ.

Given the spatial coincidence of a number of star formation tracers towards both cores \q{c1} and \q{e1}, we conclude that they are very likely to be sites of active high-mass star formation. Furthermore, there have been no UCH{\scriptsize II} regions detected towards them, as revealed by deep VLA observations \citep{Immer}. Any star formation must therefore be at a very early stage, before the high-mass star has \q{switched on}. {\textcolor{black}{This does not necessarily mean that we are observing the very initial stages of high mass star formation, as there is a significant amount of evolution involved in the formation and growth of proto-stellar cores, and the subsequent formation of high-mass stars, that occurs prior to the onset of UCH{\scriptsize II} region formation \citep[see review by][and references therein]{churchwell02}. The exact mechanism by which any embedded high-mass stars may be forming within these cores may also reduce the significance of UCH{\scriptsize II} regions as sign-posts of high-mass star formation. Given the masses of these cores, along with the substantial mass reservoirs of their natal clouds, it's likely that high accretion rates may be involved in their growth. In the presence of very high accretion rates (e.g. \textgreater \ 10$^{-3}$~M$_{\odot}$~yr$^{-1}$), the star can rapidly bloat to R \textgreater \ 100~R$_{\odot}$ \citep[e.g.][]{bloated_stars}. This ultimately delays the production of UCH{\scriptsize II} regions, and thus their non-detection towards these cores does not necessarily imply that they are very young. Nonetheless, the cores are ideal candidates for containing the early, largely-unperturbed conditions from which high-mass stars may form/be forming.}} Targeting these sources at higher spatial resolution and with line surveys will provide valuable insight into how high-mass stars form in the CMZ. We currently have multiple on-going ALMA projects to follow-up these cores along with others presented in this paper at much higher spatial resolution (\textless \ 0.1$^{\prime}$$^{\prime}$, Walker et al., in prep.).

\subsection{Do the properties of high-mass cores vary with environment?}\

%___________________________________________________________________
\begin{figure*}
\begin{center} 
\includegraphics[scale=0.88, angle=-90]{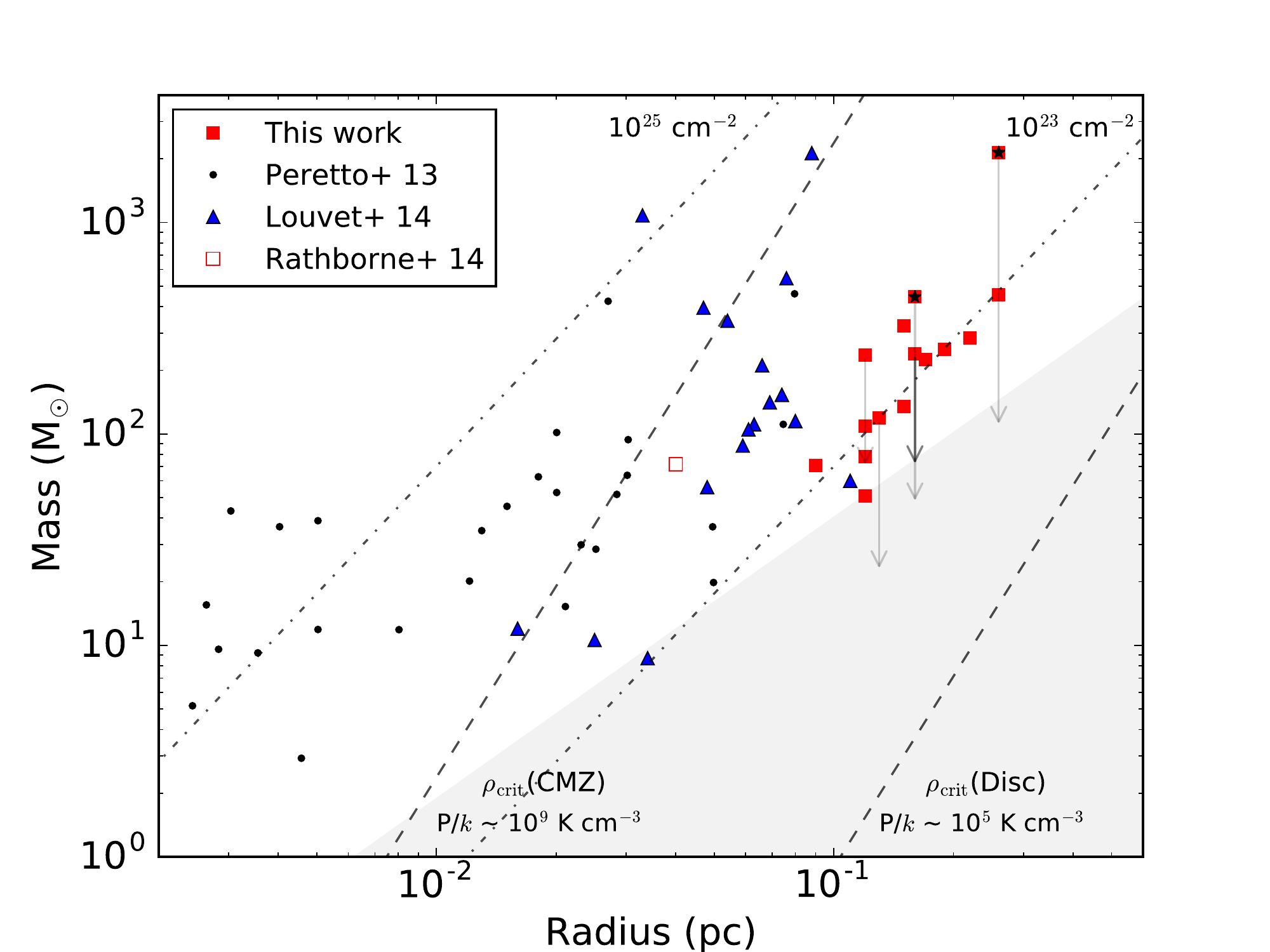}
 \caption{Mass-radius plot for all of the cores reported in our SMA sample. The solid red squares correspond to masses estimated assuming a dust temperature of 20~K and arrows indicate the possible range of mass given strong lower-mass limits estimated assuming (where possible) that the dust temperature is the same as the measured gas temperature. The red points with black star markers indicate that these cores are star-forming. Black points correspond to high-mass proto-stellar cores in the Galactic disc taken from \citet{Per_cores} and blue triangles to those from \citet{Louvet14}. The open red square corresponds to the star-forming core in cloud \q{a} (aka \q{the brick}) as seen with ALMA observations \citep{Jill_pdf_2014}. Dash/dot lines show constant column density. Dashed lines show the predicted critical volume density thresholds for both the CMZ and the Galactic disc, assuming pressures of $P/k$ = 10$^{9}$ and 10$^{5}$~K~cm$^{-3}$, respectively. The grey shaded region corresponds to the empirical massive star formation threshold proposed by \citet{Jens_10}.}
\end{center}
\end{figure*}
%%___________________________________________________________________

%___________________________________________________________________
\begin{figure*}
\begin{center}
\includegraphics[scale=0.88, angle=-90]{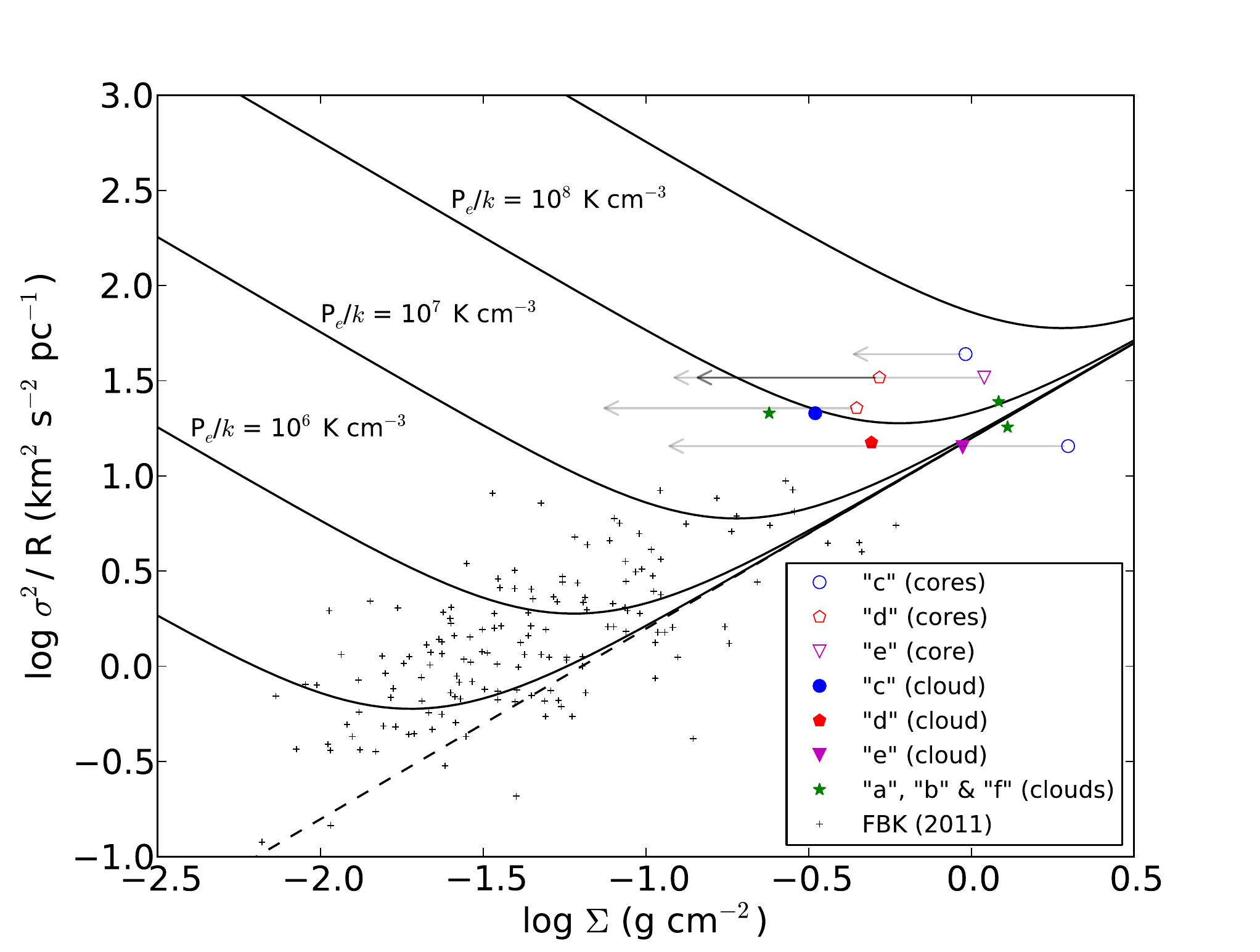}
 \caption{Comparison of the dust ridge clouds and their embedded cores with the GRS cloud sample. Black crosses show the GRS clouds as reported in \citet{Keto_pressure} (original data are from \citealt{GRS}). Solid markers indicate the different dust ridge clouds, while the open markers represent the core(s) associated with the solid markers (i.e. their parent clouds). Arrows represent the range in surface densities given our upper and lower mass limits. Curved black lines are those of constant external pressure, while the dashed line is for P$_{e}$ = 0.}
\end{center}
\end{figure*}
%%___________________________________________________________________

In $\S$1 we highlighted that these high-mass CMZ clouds are vastly under-producing stars when compared to clouds with similar properties in the less extreme environment of the Galactic disc. Recent theoretical models indicate that the CMZ environment drives high turbulent pressure in CMZ clouds. These models predict that the CMZ is undergoing episodic cycles of bursty star formation and quiescence \citep{Diederik_cmz_sf, krumholz15, krumholz16}. This heightened turbulent energy density can be seen observationally, evidenced by the comparatively large line-widths of \ap 10 -- 20~km~s$^{-1}$ seen towards these clouds on parsec scales \citep{gc_lw, Walker15, Henshaw_cmz}. So we know that on large spatial scales, the gas in the CMZ looks different to that in the Galactic disc in that it is generally orders of magnitude more dense and typical line-widths are an order of magnitude larger -- similar to the conditions seen in high-redshift galaxies \citep{Diederik_highz}. But how do the properties of the small-scale, dense structures (i.e. cores) compare to those in the disc of the Galaxy?

To compare the properties of cores in the Galactic disc and centre, we take the sample of high-mass proto-stellar cores in \citet{Per_cores}, who compiled a list of masses and radii for a sample of high-mass proto-stellar cores in the Galactic disc (Figure 6 in \citealt{Per_cores}). We also take the sample of high-mass proto-stellar cores given in Table 2 of \citet{Louvet14} for their data on the W43--MM1 ridge, which is a likely precursor to a \q{starburst cluster}, along with those for several other regions present in the literature. We scale the masses of the cores such that they are consistent with the spectral index of $\beta$ = 1.75 that we have used for mass estimates in our analysis. 

In Fig. 3 we plot the data from our sample of CMZ cores and the cores from disc clouds. The plot shows that the high-mass cores in the CMZ fit within the mass-radius relation of those in the disc reasonably well -- i.e. they are not distinctly separated. However, the cores in our sample are best described by lower volume densities of a few $\times$10$^{5}$ -- 10$^{6}$~cm$^{-3}$, while many of the disc sources are at higher volume densities. We believe that this is likely an effect of the spatial resolution of our SMA data, and that more compact fragments in these cores could approach the densities seen in regions of high mass star formation. Initial inspection of on-going ALMA observations of some of these dust ridge cores indicate that this is indeed the case (Walker et al. in prep.). The grey shaded region in Fig. 3 corresponds to the empirical massive star formation threshold that was proposed by \citet{Jens_10}. Based upon a study of a sample of nearby Galactic molecular clouds, they determine an approximate threshold for massive star formation to occur as -- $M(R)$ $\gtrsim$ $870 \textrm{M}_{\odot} \times (R/\textrm{pc})^{1.33}$. We find that all of the sources are at or above this threshold, which in the context the proposed scenario would suggest that they all have the capacity to form high-mass stars. This is certainly true of the \citet{Per_cores} and \citet{Louvet14} samples, but it currently unclear as to whether this is true for the CMZ cores. As previously discussed, only two of the cores (\q{c1} and \q{e1}) show signs of massive star formation. This difference in the star forming potential of the cores in the disc and the CMZ may be linked to an environmental dependence of the star formation process, and given that the \citet{Jens_10} threshold was derived for Solar neighbourhood clouds, it is not clear that it should hold in the CMZ.

To further investigate the effect of the different environmental conditions, we look at the CMZ cores in the context of pressure confinement. To do this, we follow the analysis of \citet{Keto_pressure}, who take the sample of clouds from the Galactic Ring Survey (GRS, \citealt{Jackson_GRS}) and study them in the context of the virial theorem for a self-gravitating isothermal spherical cloud that is subjected to a uniform external pressure, $P_{e}$. They note that based upon analysis by \citet{GRS}, the clouds are not consistent with simple virial equilibrium. They conclude that this is corrected when accounting for external pressures ranging from $P_{e}/k$ \ap 10$^{4}$ -- 10$^{6}$~K~cm$^{-3}$. We note that \citet{GRS} calculated the properties of the GRS sample using rough boxes. Analysis of the GRS clouds by \citet{Roman10} uses defined contours. This more defined approach results in the estimated virial ratios being \textless \ 1 for most of the clouds, suggesting that they may be gravitationally bound without the need for external pressure.

In Fig. 4 we replicate the plot given in Figure 3 of \citet{Keto_pressure} and over-plot both the dust ridge clouds and their embedded cores that we have detected with the SMA. The relevant parameters for plotting the dust ridge clouds can be found in Table 1, and those for the cores in Table 3. The GRS data points \citep{GRS} are shown as black crosses. The dust ridge clouds are shown as solid markers, and the cores are shown as open markers, with their colours corresponding to the clouds in which they are embedded. Note that there are fewer SMA cores displayed in this figure as it requires a measure of the line-width, which was not possible towards all of the dust cores. The dashed black line represents simple virial equilibrium, with no external pressure. The curved lines represent pressure-bounded equilibrium. These lines are described by the following equation, which is a reformulation of the pressure-bounded virial equation --

\begin{equation}
V_{0}^2 = \frac{\sigma^2}{R} = \frac{1}{3}\left(\pi \Gamma G \Sigma + \frac{4P_{e}}{\Sigma}\right)
\end{equation}
\smallskip

\noindent where $V_{0}$ is the size-linewidth scaling coefficient, $\sigma$ is the velocity dispersion, R is the radius, $\Gamma$ is related to the density structure \citep{Elme_gamma, Keto_pressure}, $\Sigma$ is the mass surface density and $P_{e}$ is the external pressure. Here we assume $\Gamma$ = 0.73, which corresponds to a centrally-concentrated density structure. This is likely valid for the SMA dust cores, but less so for the dust ridge clouds on larger scales. In \citet{Walker15, Walker16} we showed that these clouds display relatively flat surface density profiles as a result of their hierarchical internal structure.

Fig. 4 shows that, under the assumption that these cores and clouds are in pressure equilibrium, the external pressures in the CMZ would have to be of order $P_{e}/k$ \ap 10$^{8}$~K~cm$^{-3}$, which is 2--3 orders of magnitude greater than necessary for the GRS clouds in the Galactic disc. This is consistent with the observed difference in large-scale turbulent pressure between the disc and the CMZ, which for typical conditions in these environments is $P/k$ \ap 10$^{5}$~K~cm$^{-3}$ and 10$^{7-9}$~K~cm$^{-3}$, respectively \citep{Diederik_highz, Jill_pdf_2014, ymc_steve_14}. The turbulent pressure in the CMZ is therefore sufficiently high that the cores may be in equilibrium.

Comparing Figs. 3 and 4, we see that despite such high external pressures, the embedded cores in the CMZ have comparatively low densities, relative to the high-mass proto-stellar cores in the Galactic disc. Also recall our previous discussion highlighting that these cores generally lack signs of on-going star formation. Out of 15 detected sources, only 2 show signs of ongoing star formation, despite being two orders of magnitude more dense than the volume density threshold proposed by \citet{Lada10} and under pressures that are several orders of magnitude greater than in the disc. We conclude that this is further evidence for star formation being inhibited in the CMZ as a result of the high turbulent energy density, which drives up the critical volume density threshold for star formation in this environment.

\bigskip

The role of the high turbulent pressure in driving the low star formation rate in the CMZ is illustrated by the critical density lines in Figure 3 (dashed lines), which indicate the density thresholds for star formation in the models of \citet{krumholz05} and \citet{Padoan11} under the pressures seen in the Galactic disc and in the CMZ. We estimate these critical density thresholds in the context of these models as --

\begin{equation}
\rho_{\mathrm{crit}} = \frac{4\pi}{3}\frac{\alpha_{\mathrm{vir}}P_{\mathrm{turb}}\mu m_{\mathrm{H}}}{k_{B} T}
\end{equation}
\smallskip

\noindent where $\alpha_{\mathrm{vir}}$ is the virial ratio, $P_{\mathrm{turb}}$ is the turbulent pressure, $\mu$ is the molecular weight, $m_{H}$ is the Hydrogen mass, $k_{B}$ is the Boltzmann constant and T is the gas temperature.

To estimate $\rho_{\mathrm{crit}}$ in the CMZ and the disc, we assume that $P_{\mathrm{turb}}/k$ = 10$^{9}$ \& 10$^{5}$~K~cm$^{-3}$ and T = 75 \& 20~K, respectively. In both cases we assume that $\alpha_{\mathrm{vir}}$ = 1.

 The resultant critical density thresholds differ by several orders of magnitude. While the solar neighbourhood cores are all above their corresponding threshold for star formation, all of the CMZ cores in our sample are currently below the threshold appropriate for CMZ conditions. This suggests that these cores are indeed still coupled to the turbulent flow and have not been able to become self-gravitating (as found in Figure 4), thus inhibiting star formation (as concluded by \citealt{Diederik_cmz_sf} and \citealt{Jill_pdf_2014}). However, we see that the high-mass core in dust ridge cloud G0.253+0.016 (as seen with ALMA, \citealt{Jill_pdf_2014}) is the only CMZ core that lies close to the critical density threshold. This may indicate that the comparatively low densities estimated with the SMA data may be partially due to the limited spatial resolution. We also note that, at least in the case of G0.253+0.016, the magnetic field across the cloud has been shown to display little perturbation \citep[i.e. it is highly ordered,][]{Pillai_brick_bfield}. If the same is true for all of the dust ridge clouds, then it is possible that the magnetic pressure could be of the same order as that due to turbulence and may therefore be important in supporting the clouds against collapse and hence driving up the critical density threshold for star formation. Though, given the expected short lifetimes of clouds in the CMZ of \ap 1 -- 2 Myr \citep[1 --3 dynamical times;][]{Diederik_orbit, Barnes17}, it is not clear the magnetic field would have a significant effect on such short timescales.

\section{Conclusions}\ \label{sec:Conclusion}

We report our observations of the Galactic centre dust ridge clouds \q{b -- f} with the SMA at 1.3~mm. We detect a total of 15 individual dust cores above the 5$\sigma$ level, many of which are new detections. We find that the masses of these cores are substantial -- ranging \ap 50 -- 2150~M$_{\odot}$ within radii of \ap 0.1 -- 0.25~pc, with volume densities \ap 10$^{6}$~cm$^{-3}$. We find that 13 of these cores do not coincide with known star formation tracers and conclude that they represent a sample of potential pre-stellar cores that are possible precursors to high-mass stars. We detect line emission towards only 3 of these quiescent cores, and virial analysis suggests that they are unbound. However, under the high external pressures in the CMZ, we show that they may be in pressure equilibrium.

We also report the detection of two newly-discovered young, high-mass-star-forming cores -- \q{c1} and \q{e1}. These are two excellent candidates for representing the early stages of high-mass star formation in the CMZ, prior to the emission of ionising radiation.

We use the emission of H$_{2}$CO and CH$_{3}$CN to estimate the temperature of the gas in the cores, which ranges from \ap 57 -- {\textcolor{black}{220~K}}. The higher gas temperatures correspond to the cores in which we detect signs of high-mass star formation, indicating internal heating. We use the gas temperature to assume upper limits for the dust temperature and re-calculate the masses. This results in the mass estimates changing by an average of 67\%. However, we caution that these masses are very strong lower limits, as the gas is likely to be significantly warmer than the dust at these densities. This highlights the necessity of accurate dust temperature measurements on these small spatial scales in the CMZ.

The dust ridge clouds and the embedded cores are subjected to substantially higher external pressures (\ap 10$^{8}$~K~cm$^{-3}$) -- 2--3 orders of magnitude greater than typically found in the Galactic disc. Yet despite their location in such a high-pressure environment, the cores are relatively low density and only two of them show any clear signs of embedded star formation. Comparing the densities of the cores to the predicted critical density thresholds for star formation under the high pressures in the CMZ, we find that they fall below this threshold. We conclude that this is further evidence that the critical density for star formation is indeed higher in the turbulent environment of the CMZ.

\section*{Acknowledgements}

{\textcolor{black}{The authors would like to thank the anonymous referee for their feedback, which was useful in strengthening and clarifying this manuscript.}} JMDK gratefully acknowledges funding from the German Research Foundation (DFG) in the form of an Emmy Noether Research Group (grant number KR4801/1-1, PI Kruijssen) and the European Research Council (ERC) under the European Union's Horizon 2020 research and innovation programme via the ERC Starting Grant MUSTANG (grant agreement number 714907, PI Kruijssen). LCH was supported by the National Key R\&D Program of China (2016YFA0400702) and the National Science Foundation of China (11473002, 11721303).).

\bibliography{sma_bib}

\bigskip
\bigskip
\bigskip
\bigskip

\appendix
\section{\\Gas Temperature Modelling} \label{A.1}

In Section \hyperref[sec:Line]{3.2}, we highlighted the fact that we currently do not know the dust temperature on the spatial scales of our SMA resolution ($\lesssim$ 4$^{\prime\prime}$). The best constraints that we have are from Herschel observations (\ap 20~K on 33$^{\prime\prime}$ scales). We know the gas and dust in the CMZ are not thermalised on these larger scales and that the gas temperatures are significantly higher \citep[see e.g.][]{Ao_gas_temp, Brick_KJ, Adam_cmz_temp, Immer16}. However, it is possible that the gas and dust may begin to thermalise at high volume densities. \citet{Clark_brick} show this in Figure 3 of their paper, in which they model the temperature of the gas and dust as a function of volume density, showing that they begin to couple at densities \textgreater \ 10$^{6}$~cm$^{-3}$ and seem to converge at \textgreater \ 10$^{7}$~cm$^{-3}$.

Since the cores that we present in the paper are at densities of \ap 10$^{5}$ -- 10$^{6}$~cm$^{-3}$, we use molecular line emission to determine the gas temperature and reason that it may begin to couple to the dust temperature at these densities. However, we note that the densities of the cores are below those at which convergence occurs in the model of \citet{Clark_brick} and thus there may still be a significant difference between the gas and dust temperatures of up to several factors. We therefore emphasise that any masses and subsequent calculations are extreme lower limits. It is more likely that the actual dust temperature lies somewhere between these limits. In this appendix, we describe the fitting routines that we use to fit line data to estimate gas temperatures and line-widths. 

\subsection{Line fitting}\

To estimate the gas temperature via the H$_{2}$CO emission, we fit these lines under assumed LTE conditions and estimate best-fit parameters. Under the assumption of LTE, the routine constructs line profiles with a range of input parameters including peak velocities, temperatures, line-widths, and column densities, following the equations in \citet{h2co_lte}. The peak velocities of the lines can be selected manually and are fixed according to their known rest frequencies. It then minimises the differences between the constructed lines and the observed line, using the non-linear least-square fitting procedure \emph{lmfit}\footnote{http://cars9.uchicago.edu/software/python/lmfit/index.html}. The best fit is returned, along with the estimated temperature, line-width, and column density and associated errors, which we then take as the optimised fitting results\footnote{Python code for fitting 218~GHz para-H$_{2}$CO lines can be found at https://github.com/xinglunju/FFTL.}.

The primary diagnostic here is the ratio of 3$_{2,1}$-2$_{2,0}$/3$_{0,3}$-2$_{0,2}$ lines of H$_{2}$CO, where a higher fraction indicates a greater population of higher energy states and hence a higher gas temperature. The upper energy levels of the three states of para-H$_{2}$CO at \ap 218~GHz are 20.9566~K, 68.0945~K and 68.1115~K, respectively \citep[{\sc lamda} database,][]{LAMDA}. Figure A1 shows a plot of gas temperature vs. 3$_{2,1}$-2$_{2,0}$/3$_{0,3}$-2$_{0,2}$. Plotted here (dotted/dashed lines) are several non-LTE models generated using {\sc radex} \citep{Radex} for different volume densities. Also plotted (solid line) is an LTE model\footnote{The code used to generate this plot can be found at https://github.com/keflavich/h2co\_modeling/blob/5a4ce8fe9828\\4cb24ebb7b731a89012d01a965bf/examples/\\h2co\_j\%3D3\_lte\_vs\_radex.py}. This highlights the fact that temperature determination can vary significantly depending on the assumed models and volume densities. We also see that beyond certain line ratios, the models become very sensitive to small changes and are thus uncertain above certain ratios, particularly given the observational uncertainties. This is a caveat of using an LTE approximation as we have done here. Figure A1 shows that under LTE, the model becomes asymptotic beyond ratios of \ap 0.4.

%___________________________________________________________________
\begin{figure*}
\begin{center}
\includegraphics[scale=0.62, angle=-90]{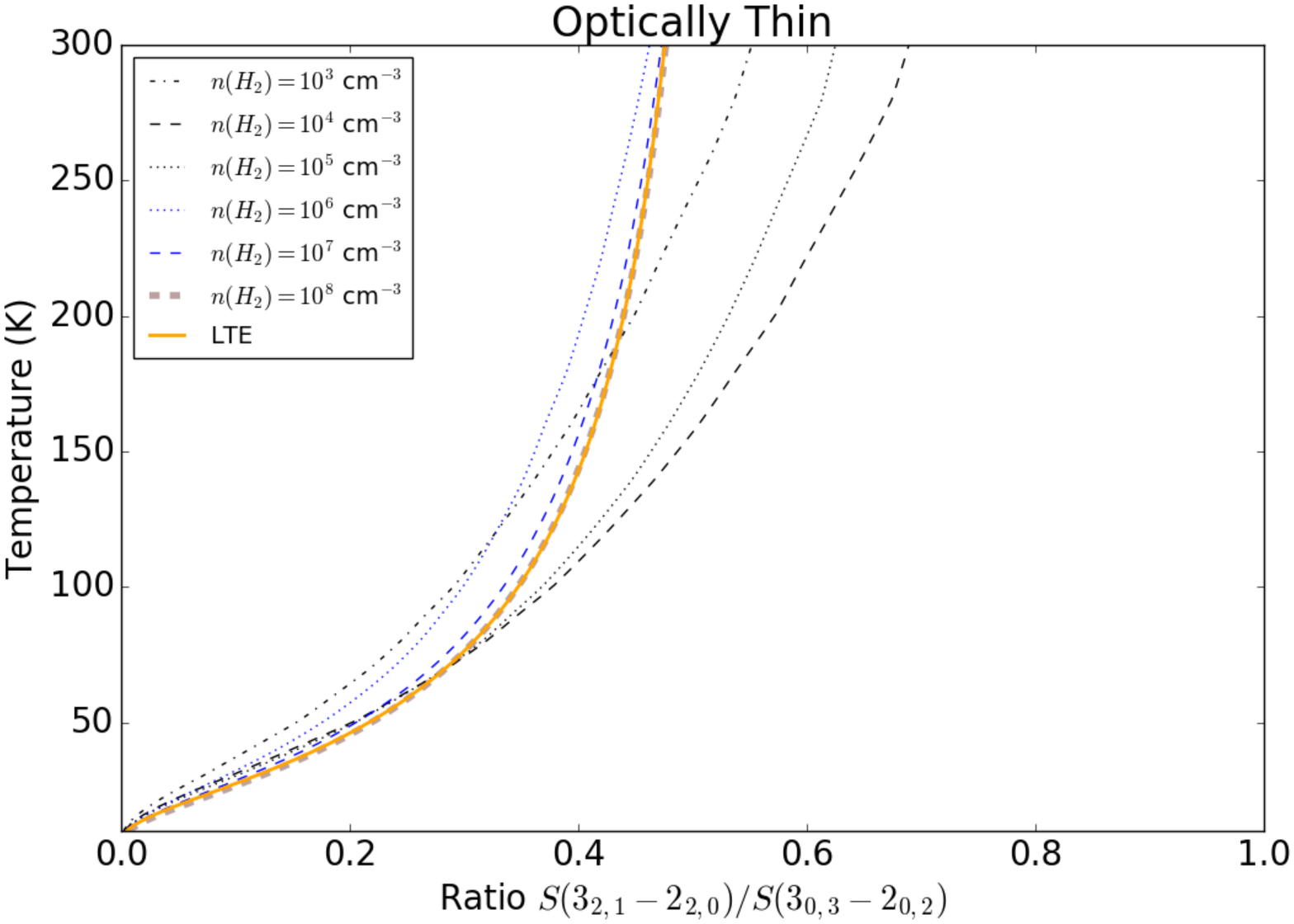}
 \caption{Gas temperature estimation as a function of the ratio of 3$_{2,1}$-2$_{2,0}$/3$_{0,3}$-2$_{0,2}$ lines of para-H$_{2}$CO for different models. The dashed/dotted lines correspond to {\sc radex} models for different volume densities and the solid line corresponds to an LTE model.} 
\end{center}
\end{figure*}
%%___________________________________________________________________

\smallskip

Of the 15 cores that we detect with the SMA, we were able to detect H$_{2}$CO above 3$\sigma$ in five of them. In the remainder of the sources, there was only very weak emission or complete non-detection. For those which we were able to fit the spectra, we then estimated the core mass, this time using the derived gas temperature as an upper limit to the dust temperature. We also report the best-fit line-width towards the cores. Using our estimated masses (both lower and upper limits) and line-widths, we then calculated the virial parameter of each core ($\alpha$ = $5R\sigma^2$/$GM$). All of these quantities are reported in Table 3. We now discuss the spectral fitting of each of these five cores.

\subsubsection{Core \q{c1}}

Our observations of dust ridge cloud \q{c} reveal a dust core that we believe contains embedded high-mass star formation. This is due to its high brightness in both the continuum and many different molecular lines, as well as its coincidence with 70~$\mu$m emission and being one of the most rich sites of rare maser emission known in the Galaxy, showing SiO, CH$_{3}$OH, H$_{2}$CO and H$_{2}$O maser emission \citep{Adam_c}. It displays strong emission in the para-H$_{2}$CO 3$_{0,3}$-2$_{0,2}$ lines at \ap 218~GHz.

Figure A2(a) shows an integrated intensity map of the para-H$_{2}$CO 3$_{0,3}$-2$_{0,2}$ line. Black contours correspond to the SMA 1.3~mm dust continuum. We see that there is good spatial coincidence between the gas and dust towards this core.

Figure A2(b) shows the para-H$_{2}$CO spectrum averaged over the spatial extent of the core, along with the best-fit and resultant temperature and line-width. While we are able to obtain good fits to the H$_{2}$CO emission of this source, we are unable to obtain a strongly-constraining fit due to the large line ratio (\ap 0.6, see Fig. A1). We therefore provide a lower limit of \ap 150 K, above which our temperature diagnostic is uncertain. This source has strong emission in all three p-H$_{2}$CO lines and in its central regions, the ratio of the 3$_{2,1}$-2$_{2,0}$/3$_{0,3}$-2$_{0,2}$ lines reaches \ap 0.8. At these ratios, there are no models that can reliably estimate the gas temperature. However, given that the core is high density ($\gtrsim$ 10$^{6}$~cm$^{-3}$), the gas temperature must be high.

{\textcolor{black}{We also detect the J=12--11 k-ladder of CH$_{3}$CN towards this core, which can also be used to estimate gas temperatures. In this core we detect the k = 0 -- 7 components of the ladder with good signal-to-noise, the upper energy levels of which are 69, 76, 97, 133, 183, 247, 326, and 419~K, respectively. It therefore allows us to trace higher temperatures more reliably than the para-H$_{2}$CO transitions. We initially attempted to fit the CH$_{3}$CN J=12-11 emission using a modified version of the LTE routine that we used to fit the H$_{2}$CO spectra, however, we found that we were not able to reliably fit the lower (k = 0 -- 3) transitions. These transitions all have very similar intensities towards this core (see Figure A3(b)), which is indicative of high optical depth. To account for the effects of opacity more thoroughly, we used XCLASS\footnote{https://xclass.astro.uni-koeln.de/} \citep[eXtended CASA Line Analysis Software Suite,][]{XCLASS} to simultaneously fit the beam-averaged CH$_{3}$CN emission along with that of its isotopologue CH$_{3}$$^{13}$CN. Figure A3(b) shows the full spectrum (black) overlaid with the best-fitted model from XCLASS (red). Overall, we are able to obtain a good fit to the data, with best fit source parameters of \ap 220~K for the temperature and \ap 2.8$\times$10$^{16}$~cm$^{-2}$ for the column density. We note that it is likely that this emission, particularly from the higher k-components, is tracing the material closer to the forming star, where the temperatures are higher and can evaporate the molecule off the dust grains and excite them to high energy states. Thus, by taking a beam-averaged spectrum, we may be averaging over different components that trace different temperatures, densities and line-widths. Despite this, we are still able to obtain a reasonable fit that demonstrates that the gas is hot and likely internally heated.}}

%We have recently obtained very high-resolution ALMA data for this source, and are in the process of performing a more detailed pixel-based modelling of the CH$_{3}$CN emission, which will allow us to probe the temperature and density structure of this core (Walker et al., in prep.).

\subsubsection{Core \q{c2}}

This core is situated \ap 0.6~pc away from \q{c1} and appears to be connected to it by a filamentary structure in the dust continuum. It is significantly less bright, with an integrated flux almost an order of magnitude smaller (see Table 3). It also does not appear to coincide with any star formation tracers and displays very little molecular line emission. It does however have H$_{2}$CO emission (Figure A4(a)), though the only line that is detected with any significance is the 3$_{0,3}$-2$_{0,2}$ line, indicating that the temperature of the gas is relatively low. Given that this is the only line detected, we place an upper limit on the gas temperature of 57.6~K, as this corresponds to the lower energy level of the 3$_{2,2}$-2$_{2,1}$ line.

\subsubsection{Cores \q{d2} \& \q{d6}}

Our SMA observations of dust ridge cloud \q{d} reveal significant dense substructure in the dust continuum, which spans the major axis of the cloud along a filamentary structure containing at least seven cores. None of these cores display any signs of star forming activity and have very weak line emission. Only cores \q{d2} and \q{d6} coincide with significant H$_{2}$CO emission (Figures A5(a, c)).

\q{d2} shows clear emission in all three para-H$_{2}$CO transitions at \ap 218~GHz, with a 3$_{2,1}$-2$_{2,0}$/3$_{0,3}$-2$_{0,2}$ ratio of \ap 0.3, which yields a corresponding fit of 86.3~K ($\pm$ 20.4~K).

\q{d6} has clear emission in the 3$_{0,3}$-2$_{0,2}$ transition, with signs of very weak emission from the other two transitions. Given that these other transitions are around the noise level, we cannot obtain a constraining fit and so we again place an upper limit on the gas temperature of 57.6~K.

\subsubsection{Core \q{e1}}

Cloud \q{e} is the most massive (1.1 -- 1.5$\times$10$^{5}$~M$_{\odot}$) cloud along the dust ridge and is seen to contain two dense dust cores in our SMA data. One of these, \q{e1}, stands out as it coincides with both H$_{2}$O and CH$_{3}$OH Class II maser emission, as well as a 70~$\mu$m source. It is therefore believed to be a potential site of young high-mass star formation. We also see that the H$_{2}$CO emission traces the dust core well (Figure A6(a)). We were unable to obtain a reliable fit to the H$_{2}$CO lines as the large line ratio of 3$_{2,1}$-2$_{2,0}$/3$_{0,3}$-2$_{0,2}$ cannot be fit by any models. We assume a lower limit of 150~K, since our LTE method cannot reliably discern between temperatures beyond this. However, given that we know that this core is dense, this large line ratio means that the gas must be hot. We note that there appears to be a slight excess in the emission of the 3$_{2,1}$-2$_{2,0}$ line. The exact cause of this is not known, but it may suggest that there are multiple components present and/or that there are significant temperature/density gradients within the core.

%___________________________________________________________________
    \begin{figure*}
        \centering
        \begin{subfigure}[b]{0.475\textwidth}
            \centering
            \includegraphics[width=\textwidth, angle=-90, scale=0.9]{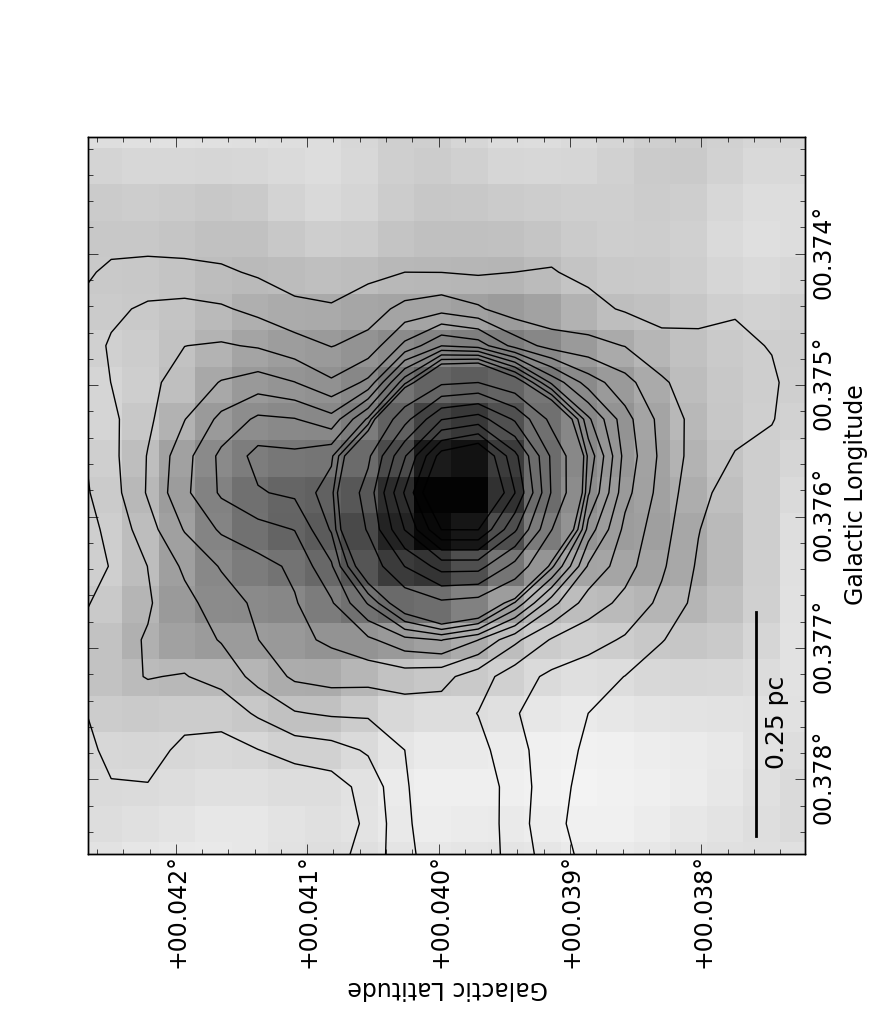}
            \caption[]%
            {{\small}}    
        \end{subfigure}
        \begin{subfigure}[b]{0.475\textwidth}   
            \centering 
			\includegraphics[width=\textwidth, angle=-90, scale=0.85]{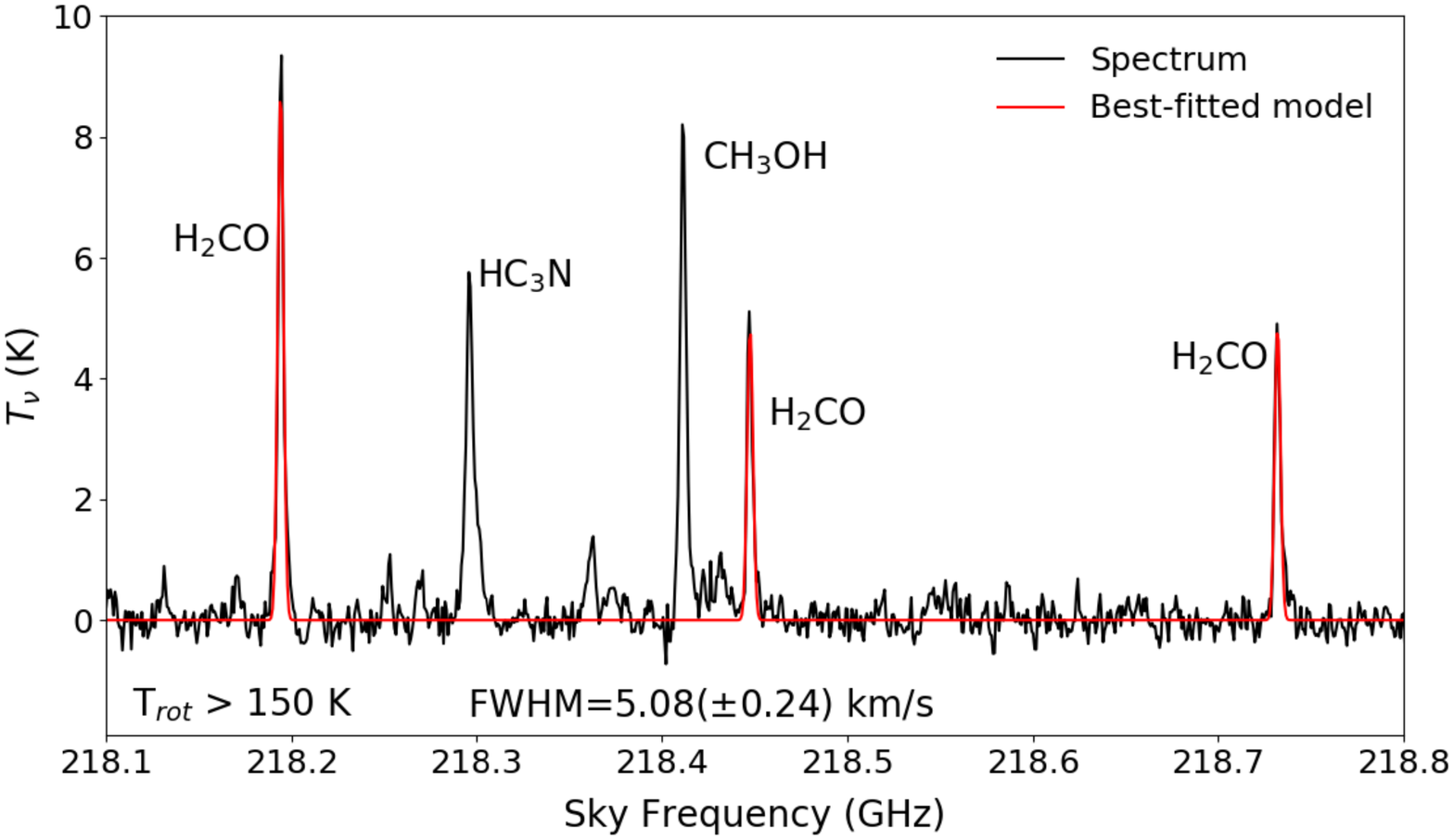}
            \caption[]%
            {{\small}}    
        \end{subfigure}

        \caption[]
        {(a) Integrated intensity map of the para-H$_{2}$CO 3$_{0,3}$-2$_{0,2}$ transition towards \q{c1}. Black contours are SMA dust continuum. (b) Spatially-averaged spectrum of the para-H$_{2}$CO lines (plus HC$_{3}$N and CH$_{3}$OH). Using an LTE line-fitting routine, we fit the H$_{2}$CO lines. Due to the large line ratios, we are unable to obtain a reliable gas temperature (see Fig. A1). We therefore provide a lower limit of 150~K, above which our temperature diagnostic is uncertain.} 
    \end{figure*}
%___________________________________________________________________

%___________________________________________________________________
    \begin{figure*}
        \centering
        \begin{subfigure}[b]{0.475\textwidth}
            \centering
            \includegraphics[width=\textwidth, angle=-90, scale=0.9]{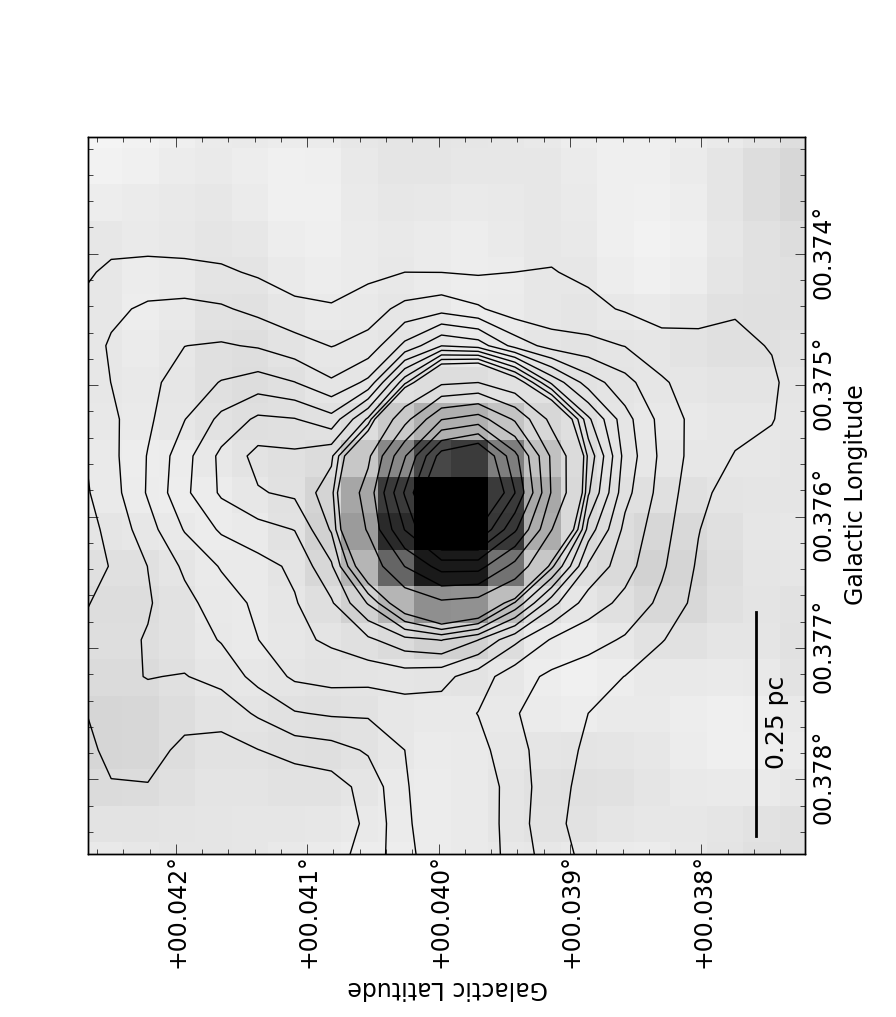}
            \caption[]%
            {{\small}}    
        \end{subfigure}
%        \vskip\baselineskip
        \begin{subfigure}[b]{0.475\textwidth}   
            \centering 
			\includegraphics[width=\textwidth, angle=-90, scale=0.65]{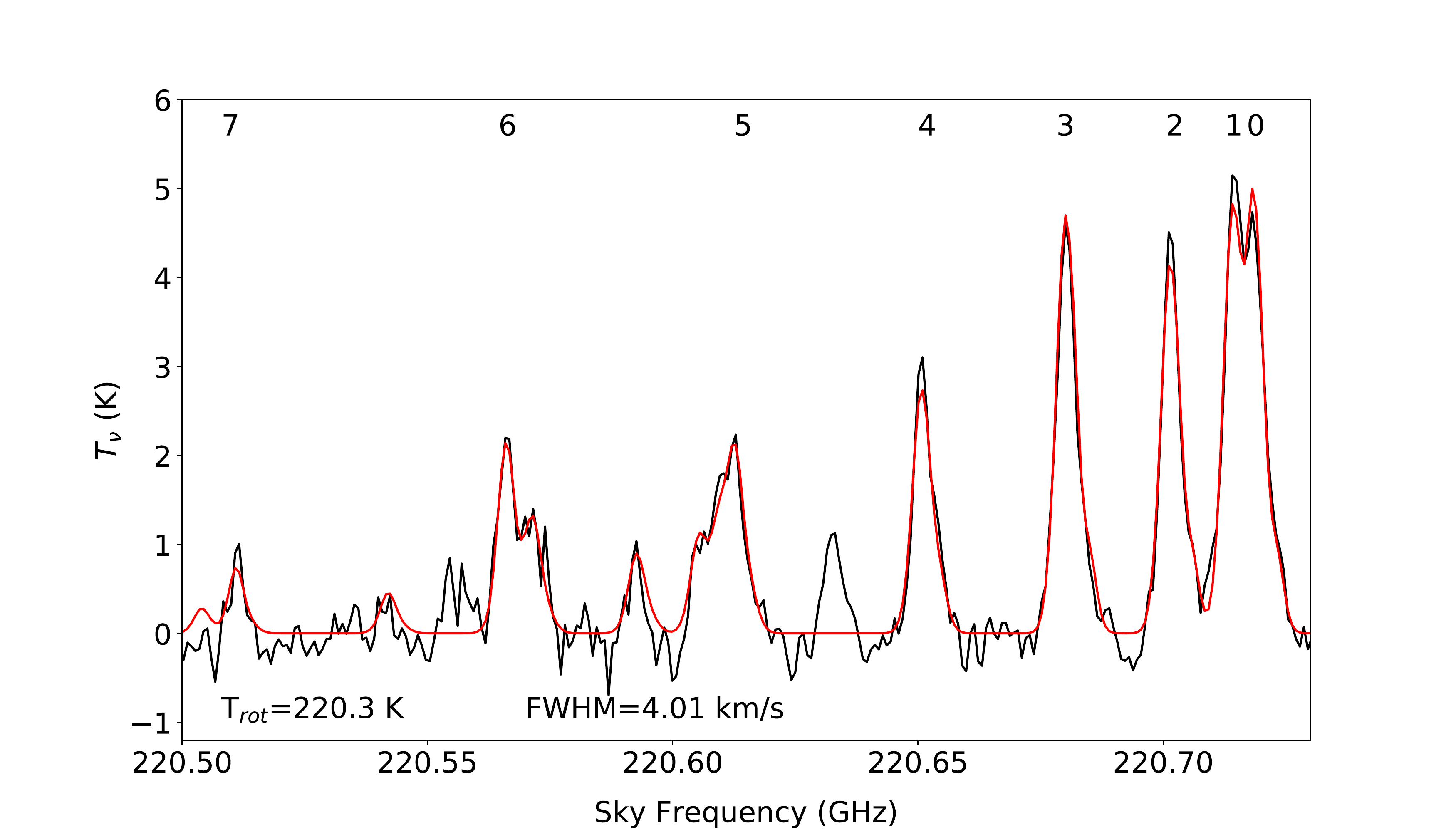}
            \caption[]%
            {{\small}}    
        \end{subfigure}
%         \quad
%        \begin{subfigure}[b]{0.475\textwidth}   
%            \centering 
%            \includegraphics[width=\textwidth, angle=-90, scale=0.8]{ch3cn_moment_map.pdf}
%            \caption[]%
%            {{\small (d)}}    
%        \end{subfigure}

        \caption[]
        {{\textcolor{black}{(a) Integrated intensity map of the CH$_{3}$CN J=12--11 k-ladder towards \q{c1}. Black contours are SMA dust continuum. (b) Spatially averaged spectrum of the CH$_{3}$CN J=12--11 k-ladder. The k = 0 -- 7 components are clearly detected (labelled above the corresponding transition), the upper energy levels of which are 69, 76, 97, 133, 183, 247, 326, and 419~K, respectively. This suggests significant internal heating, with high gas temperatures likely occurring close to the embedded proto-star(s). Modelling the spectrum with XCLASS, we estimate a beam-averaged gas temperature of 220~K.}}} 
    \end{figure*}
%___________________________________________________________________

%___________________________________________________________________
    \begin{figure*}
        \centering
        \begin{subfigure}[b]{0.475\textwidth}
            \centering
            \includegraphics[width=\textwidth, angle=-90, scale=0.9]{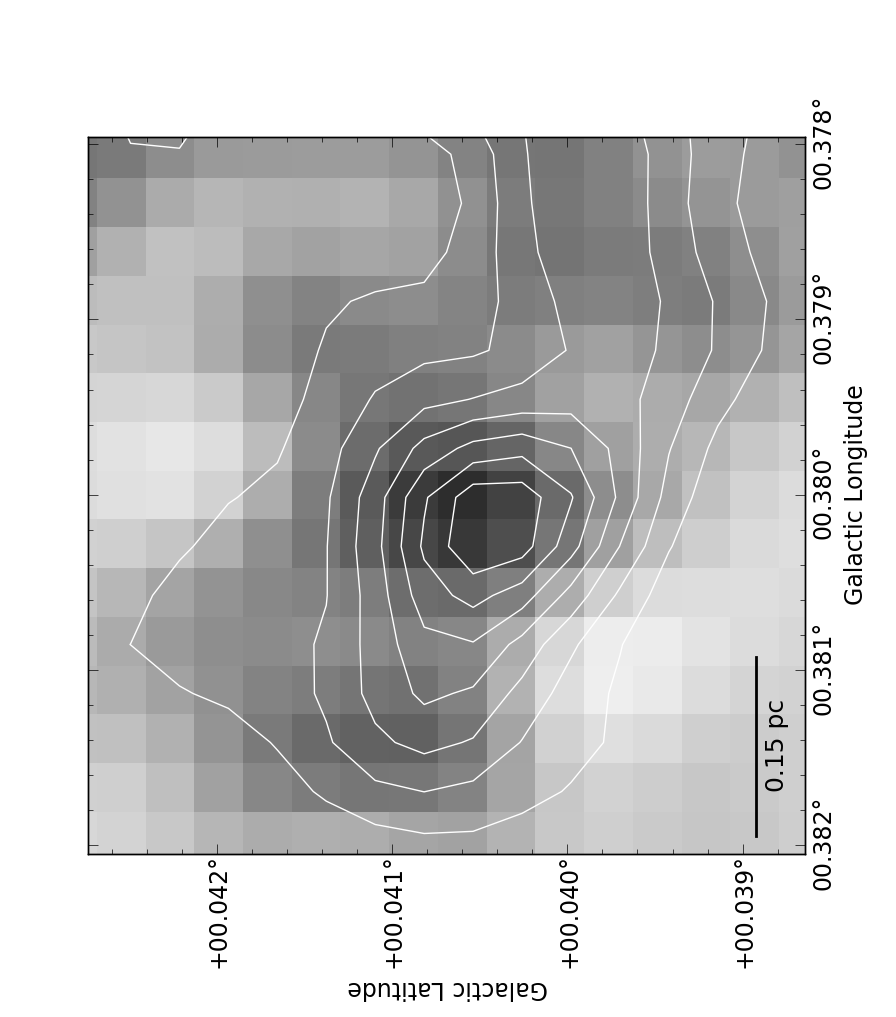}
            \caption[]%
            {{\small}}    
        \end{subfigure}
%        \hfill
        \begin{subfigure}[b]{0.475\textwidth}  
            \centering 
            \includegraphics[width=\textwidth, angle=-90, scale=0.85]{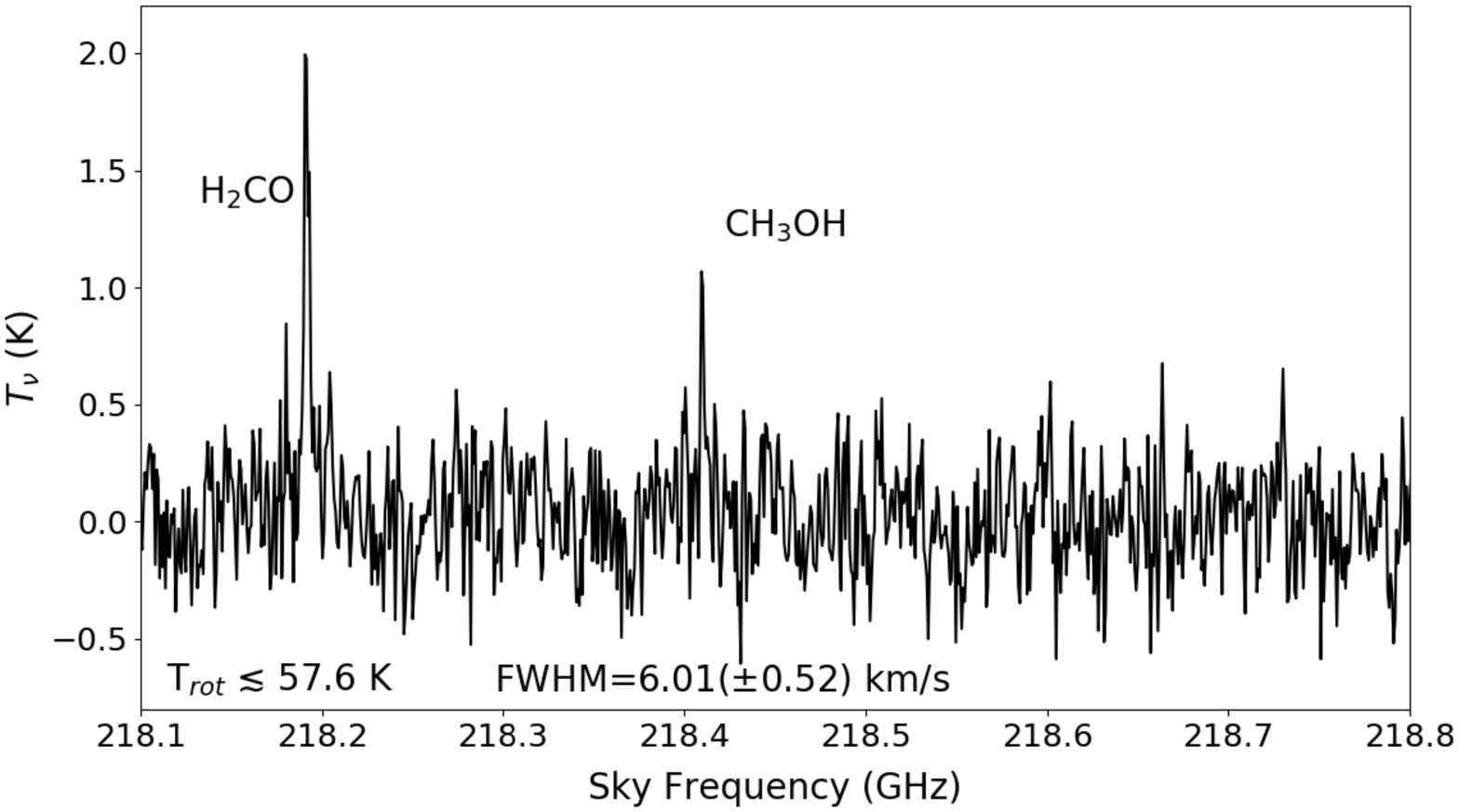}
            \caption[]%
            {{\small}}    
        \end{subfigure}
        \caption[]
        {(a) Integrated intensity map of the para-H$_{2}$CO 3$_{0,3}$-2$_{0,2}$ transition towards \q{c2}. White contours are SMA dust continuum. (b) Spatially-averaged spectrum of the para-H$_{2}$CO lines. Given that only the 3$_{0,3}$-2$_{0,2}$ line is well detected, we place an upper limit on the gas temperature of 57.6~K, as this corresponds to the lower energy level of the 3$_{2,2}$-2$_{2,1}$ line.} 
    \end{figure*}
%___________________________________________________________________

%___________________________________________________________________
    \begin{figure*}
        \centering
        \begin{subfigure}[b]{0.475\textwidth}
            \centering
            \includegraphics[width=\textwidth, angle=-90, scale=0.9]{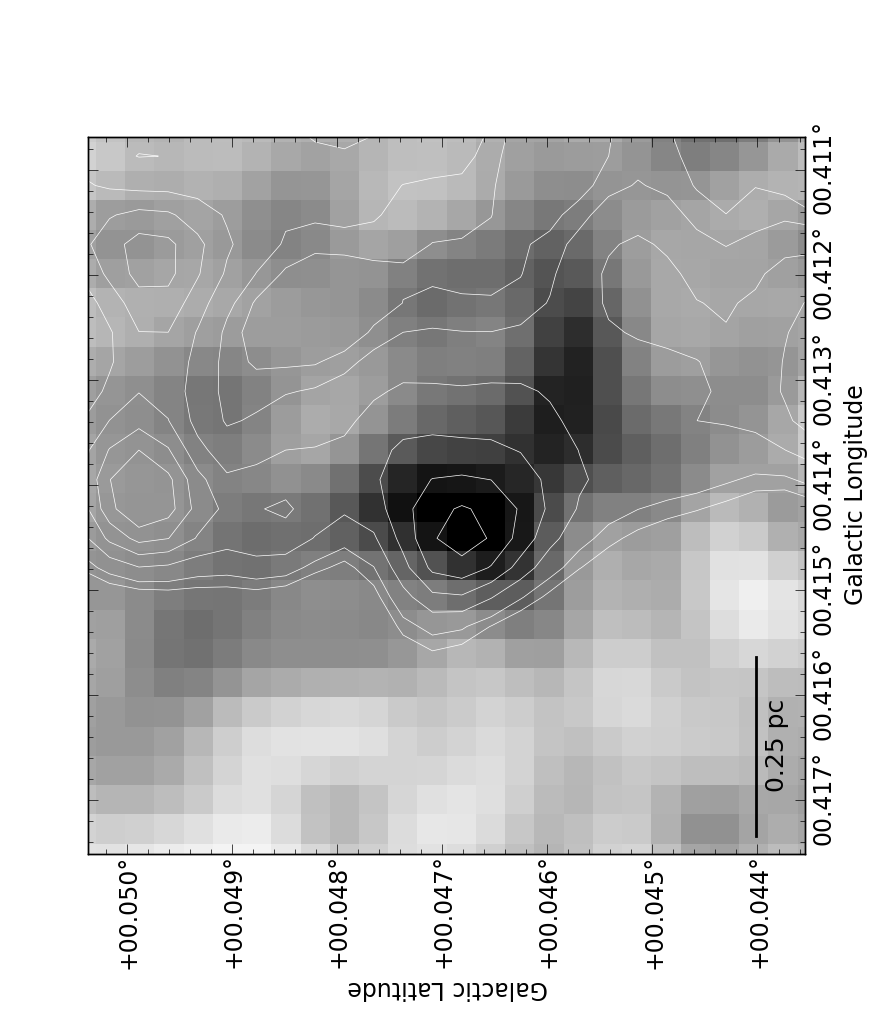}
            \caption[]%
            {{\small}}    
        \end{subfigure}
%        \hfill
        \begin{subfigure}[b]{0.475\textwidth}  
            \centering 
            \includegraphics[width=\textwidth, angle=-90, scale=0.85]{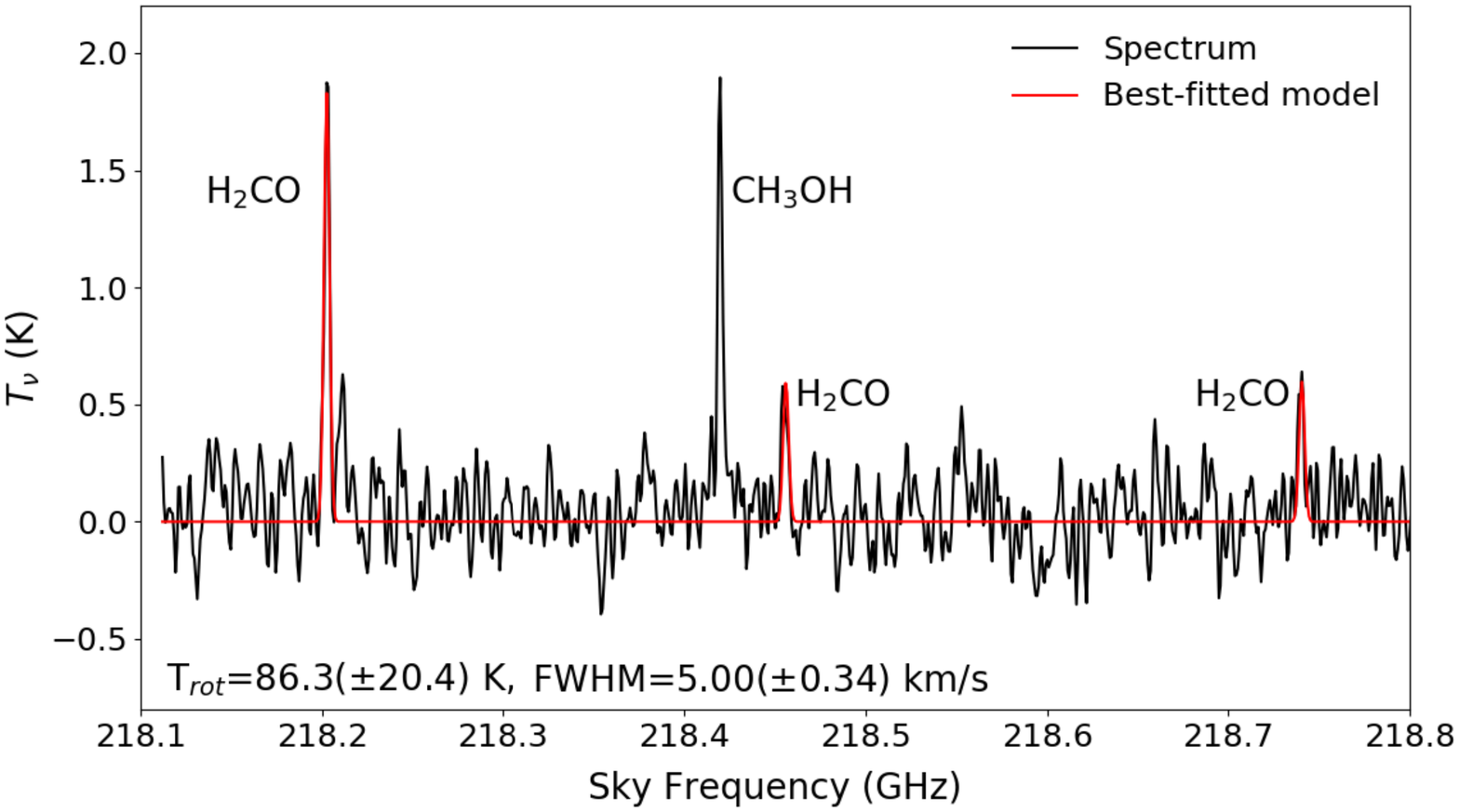}
            \caption[]%
            {{\small}}    
        \end{subfigure}
        \vskip\baselineskip
        \begin{subfigure}[b]{0.475\textwidth}   
            \centering 
            \includegraphics[width=\textwidth, angle=-90, scale=0.9]{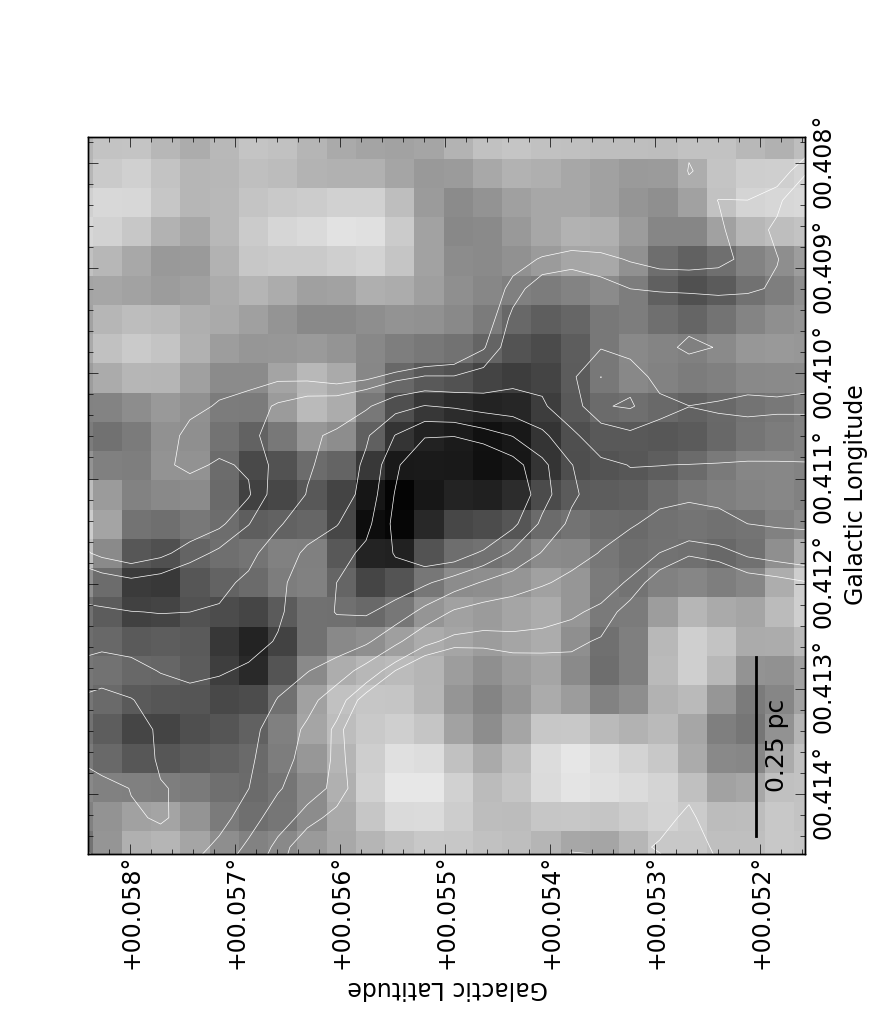}
            \caption[]%
            {{\small}}    
        \end{subfigure}
         \quad
        \begin{subfigure}[b]{0.475\textwidth}   
            \centering 
            \includegraphics[width=\textwidth, angle=-90, scale=0.8]{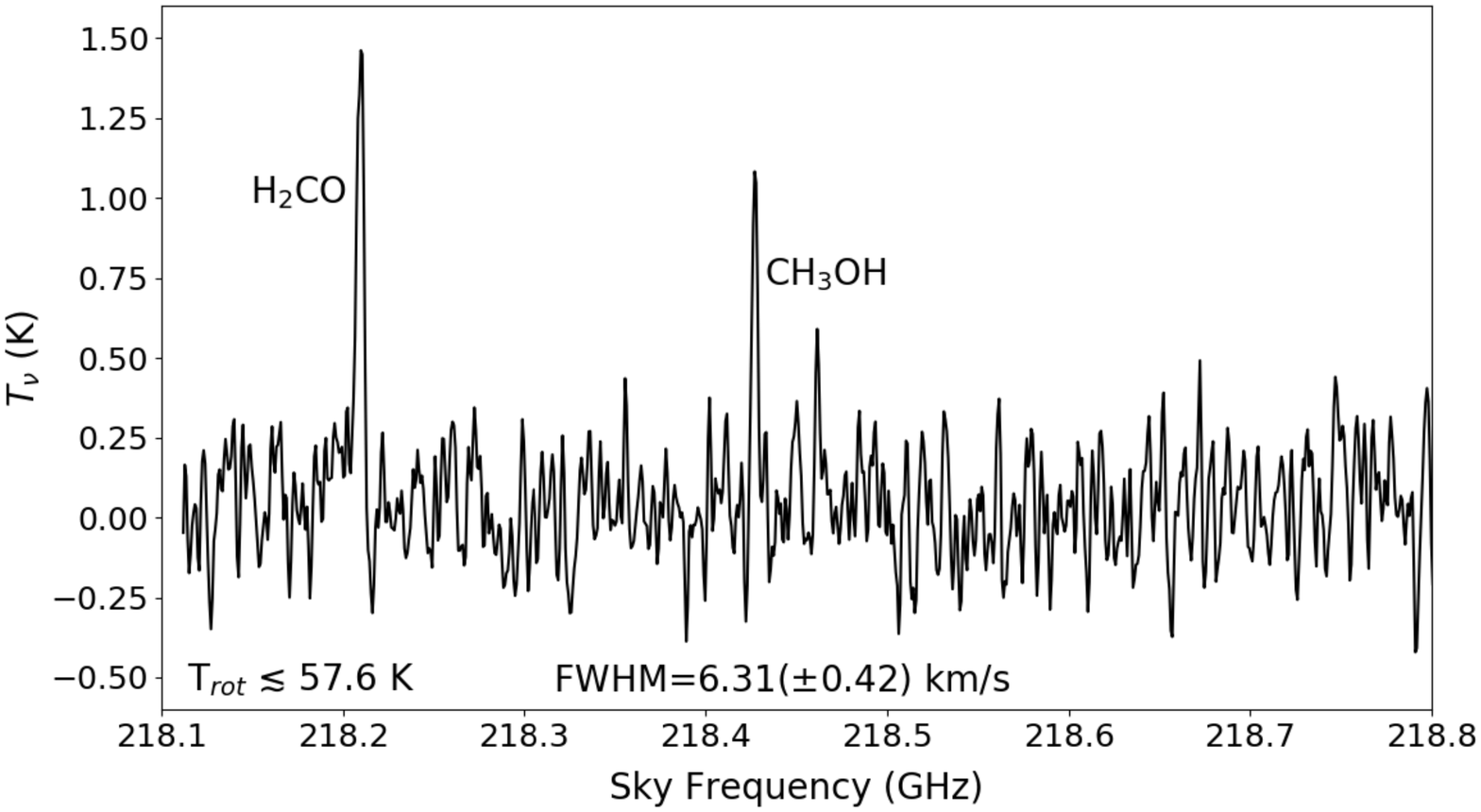}
            \caption[]%
            {{\small}}    
        \end{subfigure}

        \caption[]
        {(a) Integrated intensity map of the para-H$_{2}$CO 3$_{0,3}$-2$_{0,2}$ transition towards \q{d2}. White contours are SMA dust continuum. (b) Spatially-averaged spectrum of the para-H$_{2}$CO lines. Using an LTE line-fitting routine, we fit the H$_{2}$CO lines to derive a beam-averaged gas temperature of 86.3~K ($\pm$ 20.4~K). (c) Integrated intensity map of the para-H$_{2}$CO 3$_{0,3}$-2$_{0,2}$ transition towards \q{d6}. White contours are SMA dust continuum. (d) Spatially-averaged spectrum of the para-H$_{2}$CO lines. Given that only the 3$_{0,3}$-2$_{0,2}$ line is well detected, we place an upper limit on the gas temperature of 57.6~K, as this corresponds to the lower energy level of the 3$_{2,2}$-2$_{2,1}$ line.} 
    \end{figure*}
%___________________________________________________________________

%___________________________________________________________________
    \begin{figure*}
        \centering
        \begin{subfigure}[b]{0.475\textwidth}
            \centering
            \includegraphics[width=\textwidth, angle=-90, scale=0.9]{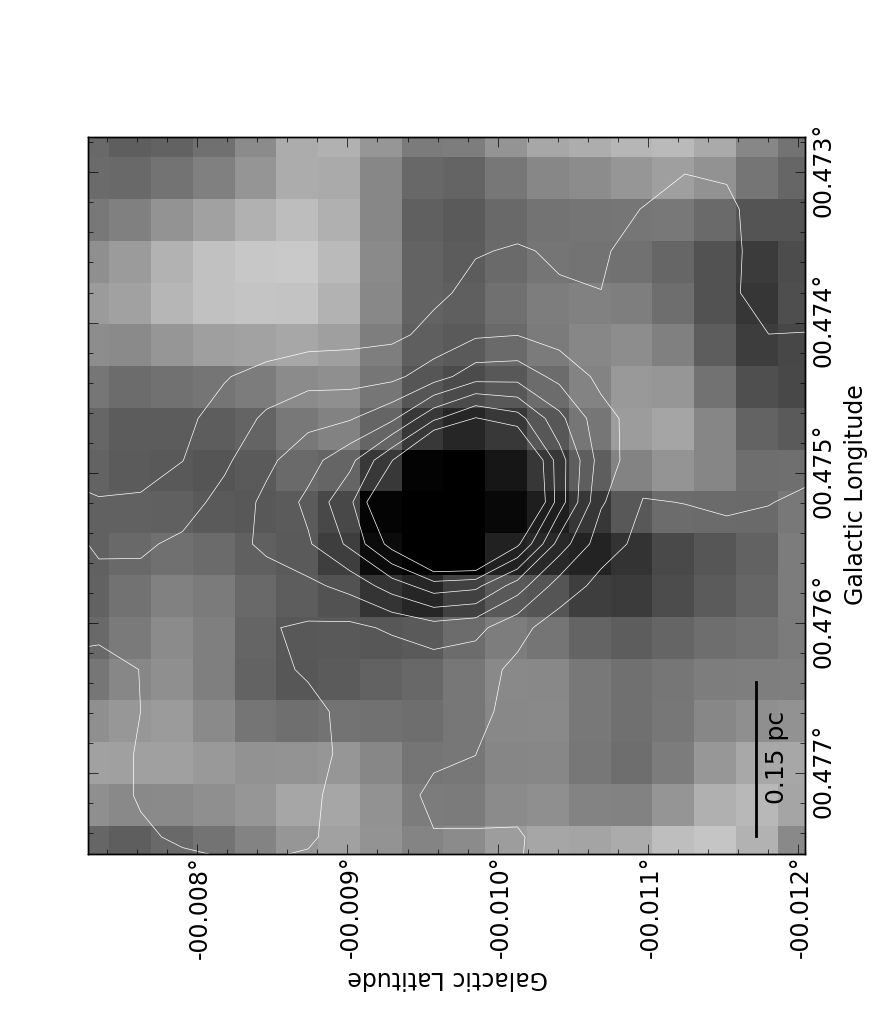}
            \caption[]%
            {{\small}}    
        \end{subfigure}
%        \hfill
        \begin{subfigure}[b]{0.475\textwidth}  
            \centering 
            \includegraphics[width=\textwidth, angle=-90, scale=0.85]{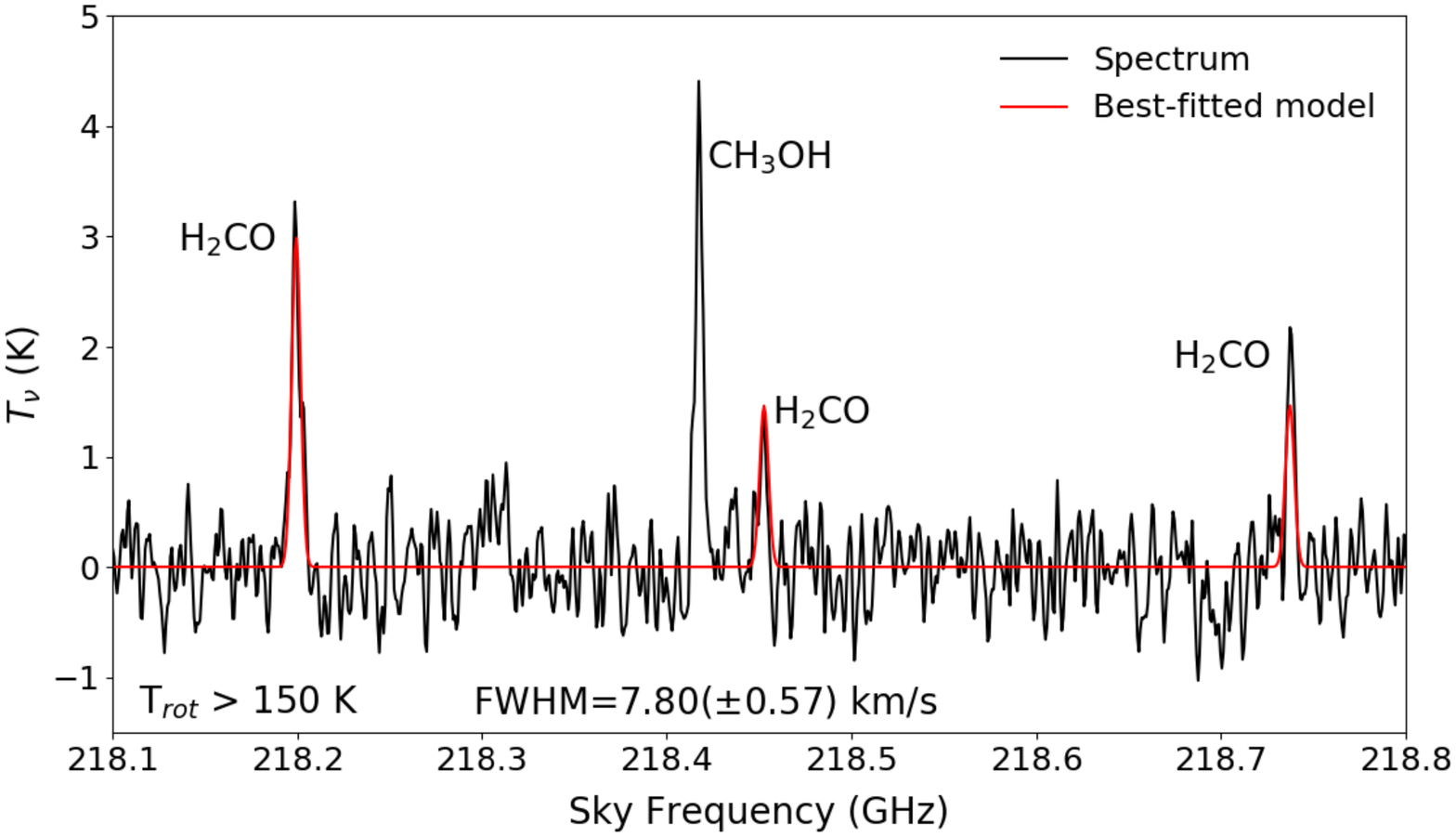}
            \caption[]%
            {{\small}}    
        \end{subfigure}
        \caption[]
        {(a) Integrated intensity map of the para-H$_{2}$CO 3$_{0,3}$-2$_{0,2}$ transition towards \q{e1}. White contours are SMA dust continuum. (b) Spatially-averaged spectrum of the para-H$_{2}$CO lines. We were unable to obtain a reliable fit to the H$_{2}$CO lines as the large line ratio of 3$_{2,1}$-2$_{2,0}$/3$_{0,3}$-2$_{0,2}$ cannot be fit by any models. We assume a lower limit of 150~K, since our LTE method cannot reliably discern between temperatures beyond this. The cause for the apparent excess in the 3$_{2,1}$-2$_{2,0}$ line is not known, but may suggest that the emission comes from more than one component, and may have a temperature/density gradient.
} 
    \end{figure*}
%___________________________________________________________________

\bsp

\label{lastpage}

\end{document}